\documentclass[a4paper,british]{amsart}
\usepackage[T1]{fontenc}
\usepackage[latin9]{inputenc}
\usepackage{float}
\usepackage{url}
\usepackage{bm}
\usepackage{amsthm}
\usepackage{amssymb}
\usepackage{graphicx}
\usepackage{esint}
\usepackage[authoryear]{natbib}

\makeatletter

\pdfpageheight\paperheight
\pdfpagewidth\paperwidth

\providecommand{\tabularnewline}{\\}
\floatstyle{ruled}
\newfloat{algorithm}{tbp}{loa}
\providecommand{\algorithmname}{Algorithm}
\floatname{algorithm}{\protect\algorithmname}

\numberwithin{equation}{section}
\numberwithin{figure}{section}

\usepackage{bbm} 
\newcommand{\ind}{\mathbbm{1}} 

\newcommand{\bx}{\bm{x}}

\newcommand{\bz}{\bm{z}}
\newcommand{\bu}{\bm{u}}
\newcommand{\bbeta}{\bm{\beta}}
\newcommand{\bsigsq}{\bm{\sigma}^2}
\newcommand{\bgamma}{\bm{\gamma}}
\newcommand{\mom}{\bm{\alpha}} 
\newcommand{\data}{\mathcal{D}}
\newcommand{\sgn}{\mathrm{sgn}}
\newcommand{\betamap}{\bbeta_{\mathrm{MAP}}}
\newcommand{\eqdef}{:=}
\newcommand{\dd}{\mathrm{d}}
\newcommand{\OO}{\mathcal{O}}
\newcommand{\ndata}{n_\data}
\newcommand{\setR}{\mathbb{R}}

\newcommand{\E}{\mathbb{E}} 
\newcommand{\R}{\mathbb{R}} 

\newcommand{\Norm}{\mathrm{N}}
\newcommand{\Gam}{\mathrm{Gamma}}
\newcommand{\InvGam}{\mathrm{Inv-Gamma}}
\newcommand{\Chisq}{\mathrm{Chi}^2}

\@ifundefined{showcaptionsetup}{}{%
 \PassOptionsToPackage{caption=false}{subfig}}
\usepackage{subfig}
\makeatother

\usepackage{babel}
\begin{document}

\author{Nicolas Chopin and James Ridgway}

\title[Leave Pima indians alone]{Leave Pima Indians alone: binary regression as a benchmark for Bayesian
computation}
\begin{abstract}
Whenever a new approach to perform Bayesian computation is introduced,
a common practice is to showcase this approach on a binary regression
model and datasets of moderate size. This paper discusses to which
extent this practice is sound. It also reviews the current state of
the art of Bayesian computation, using binary regression as a running
example. Both sampling-based algorithms (importance sampling, MCMC
and SMC) and fast approximations (Laplace and EP) are covered. Extensive
numerical results are provided, some of which might go against conventional
wisdom regarding the effectiveness of certain algorithms. Implications
for other problems (variable selection) and other models are also
discussed. 
\end{abstract}

\maketitle

\section{Introduction}

The field of Bayesian computation seems hard to track these days,
as it is blossoming in many directions. MCMC (Markov chain Monte Carlo)
remains the main approach, but it is no longer restricted to Gibbs
sampling and Hastings-Metropolis, as it includes more advanced, Physics-inspired
methods, such as HMC \citep[Hybrid Monte Carlo, ][]{Neal2010HMC}
and its variants \citep{girolami2011riemann,shahbaba2011split,NUTS}.
On the other hand, there is also a growing interest for alternatives
to MCMC, such as SMC (Sequential Monte Carlo, e.g. \citealp{DelDouJas:SMC}),
nested sampling \citep{SkillingNested}, or the fast approximations
that originated from machine learning, such as Variational Bayes \citep[e.g. ][Chap. 10]{Bishop:book},
and EP \citep[Expectation Propagation, ][]{minka2001expectation}.
Even Laplace approximation has resurfaced in particular thanks to
the INLA methodology \citep{rue2009approximate}. 

One thing however that all these approaches have in common is they
are almost always illustrated by a binary regression example; see
e.g. the aforementioned papers. In other words, binary regressions
models, such as probit or logit, are a de facto benchmark for Bayesian
computation. 

This remark leads to several questions. Are binary regression models
a reasonable benchmark for Bayesian computation? Should they be used
then to develop a `benchmark culture' in Bayesian computation, like
in e.g. optimisation? And practically, which of these methods actually
`works best' for approximating the posterior distribution of a binary
regression model? 

The objective of this paper is to answer these questions. As the ironic
title suggests, our findings shall lead to us be critical of certain
current practices. Specifically, most papers seem content with comparing
some new algorithm with Gibbs sampling, on a few small datasets, such
as the well-known Pima Indians diabetes dataset ($8$ covariates).
But we shall see that, for such datasets, approaches that are even
more basic than Gibbs sampling are actually hard to beat. In other
words, datasets considered in the literature may be too toy-like to
be used as a relevant benchmark. On the other hand, if ones considers
larger datasets (with say 100 covariates), then not so many approaches
seem to remain competitive. 

We would also like to discuss \emph{how} Bayesian computation algorithms
should be compared. One obvious criterion is the error versus CPU
time trade-off; this implies discussing which posterior quantities
one may be need to approximate. A related point is whether the considered
method comes with a simple way to evaluate the numerical error. Other
criteria of interest are: (a) how easy to implement is the considered
method? (b) how generic is it? (does changing the prior or the link
function require a complete rewrite of the source code?) (c) to which
extent does it require manual tuning to obtain good performances?
(d) is it amenable to parallelisation? Points (a) and (b) are rarely
discussed in Statistics, but relate to the important fact that, the
simpler the program, the easier it is to maintain, and to make it
bug-free. Regarding point (c), we warn beforehand that, as a matter
of principle, we shall refuse to manually tune an algorithm on a per
dataset basis. Rather, we will discuss, for each approach, some (hopefully
reasonable) general recipe for how to choose the tuning parameters.
This has two motivations. First, human time is far more valuable that
computer time: \citet{Cook:time} mentions that one hour of CPU time
is today three orders of magnitude less expensive than one hour of
pay for a programmer (or similarly a scientist). Second, any method
requiring too much manual tuning through trial and error may be practically
of no use beyond a small number of experts. 

Finally, we also hope this paper may serve as an up to date review
of the state of Bayesian computation. We believe this review to be
timely for a number of reasons. First, as already mentioned, because
Bayesian computation seems to develop currently in several different
directions. Second, and this relates to criterion (d), the current
interest in parallel computation \citep{Lee2010,Suchard2010} may
require a re-assessment of Bayesian computational methods: method
A may perform better than method B on a single core architecture,
while performing much worse on a parallel architecture. Finally, although
the phrase `big data' seems to be a tired trope already, it is certainly
true that datasets are getting bigger and bigger, which in return
means that statistical methods needs to be evaluated on bigger and
bigger datasets. To be fair, we will not really consider in this work
the kind of huge datasets that pertain to `big data', but we will
at least strive to move away from the kind of `ridiculously small'
data encountered too often in Bayesian computation papers.

The paper is structured as follows. Section \ref{sec:Binary-regression-models}
covers certain useful preliminaries on binary regression models. Section
\ref{sec:Fast-approximation-methods} discusses fast approximations,
that is, deterministic algorithms that offer an approximation of the
posterior, at a lower cost than sampling-based methods. Section \ref{sec:Exact-methods}
discusses `exact', sampling-based methods. Section \ref{sec:Numerics}
is the most important part of the paper, as it contains an extensive
numerical comparison of all these methods. Section \ref{sec:Variable-selection}
discusses variable selection. Section \ref{sec:Conclusion-and-extensions}
discusses our findings, and their implications for both end users
and Bayesian computation experts.

\section{Preliminaries: binary regression models\label{sec:Binary-regression-models}}

\subsection{Likelihood, prior\label{sub:Likelihood,-prior}}

The likelihood of a binary regression model have the generic expression
\begin{equation}
p(\data|\bm{\beta})=\prod_{i=1}^{\ndata}F(y_{i}\bm{\beta}^{T}\bx_{i})\label{eq:lik}
\end{equation}
where the data $\data$ consist of $n$ responses $y_{i}\in\left\{ -1,1\right\} $
and $n$ vectors $\bx_{i}$ of $p$ covariates, and $F$ is some CDF
(cumulative distribution function) that transforms the linear form
$y_{i}\bm{\beta}^{T}\bx_{i}$ into a probability. Taking $F=\Phi$,
the standard normal CDF, gives the probit model, while taking $F=L$,
the logistic CDF, $L(x)=1/\left(1+e^{-x}\right)$, leads to the logistic
model. Other choices could be considered, such as e.g. the CDF of
a Student distribution (robit model) to better accommodate outliers. 

We follow \citet{MR2655663}'s recommendation to standardise the predictors
in a preliminary step: non-binary predictors have mean $0$ and standard
deviation $0.5$, binary predictors have mean $0$ and range $1$,
and the intercept (if present) is set to $1$. This standardisation
facilitates prior specification: one then may set up a ``weakly informative''
prior for $\bm{\beta}$, that is a proper prior that assigns a low
probability that the marginal effect of one predictor is outside a
reasonable range. Specifically, we shall consider two priors $p(\bbeta)$
in this work: (a) the default prior recommended by \citet{MR2655663},
a product of independent Cauchys with centre 0 and scale $10$ for
the constant predictor, $2.5$ for all the other predictors (henceforth,
the Cauchy prior); and (b) a product of independent Gaussians with
mean $0$ and standard deviation equal to twice the scale of the Cauchy
prior (henceforth the Gaussian prior). 

Of course, other priors could be considered, such as e.g. Jeffreys'
prior \citep{firth1993bias}, or a Laplace prior \citep{kaban2007bayesian}.
Our main point in considering the two priors above is to determine
to which extent certain Bayesian computation methods may be prior-dependent,
either in their implementation (e.g. Gibbs sampling) or in their performance,
or both. In particular, one may expect the Cauchy prior to be more
difficult to deal with, given its heavy tails.

\subsection{Posterior maximisation (Gaussian prior)\label{sub:Posterior-max-Gauss-prior}}

We explain in this section how to quickly compute the mode, and the
Hessian at the mode, of the posterior: 
\[
p(\bbeta|\data)=\frac{p(\bbeta)p(\data|\bbeta)}{p(\data)},\quad p(\data)=\int_{\R^{d}}p(\bbeta)p(\data|\bbeta)\,d\bbeta,
\]
where $p(\bbeta)$ is one of the two priors presented in the previous
section, and $Z(\data)$ is the marginal likelihood of the data (also
known as the evidence). These quantities will prove useful later,
in particular to tune certain of the considered methods. 

The two first derivatives of the log-posterior density may be computed
as:
\begin{align*}
\frac{\partial}{\partial\bbeta}\log p(\bbeta|\data) & =\frac{\partial}{\partial\bbeta}\log p(\bbeta)+\frac{\partial}{\partial\bbeta}\log p(\data|\bbeta),\\
\quad\frac{\partial^{2}}{\partial\bbeta\partial\bbeta^{T}}\log p(\bbeta|\data) & =\frac{\partial^{2}}{\partial\bbeta\partial\bbeta^{T}}\log p(\bbeta)+\frac{\partial^{2}}{\partial\bbeta\partial\bbeta^{T}}\log p(\data|\bbeta)
\end{align*}
where
\begin{align*}
\frac{\partial}{\partial\bbeta}\log p(\data|\bbeta) & =\sum_{i=1}^{\ndata}\left(\log F\right)'(y_{i}\bbeta^{T}\bx_{i})y_{i}\bx_{i}\\
\frac{\partial^{2}}{\partial\bbeta\partial\bbeta^{T}}\log p(\data|\bbeta) & =\sum_{i=1}^{\ndata}\left(\log F\right)''(y_{i}\bbeta^{T}\bx_{i})\bx_{i}\bx_{i}^{T}
\end{align*}
and $(\log F)'$ and $\left(\log F\right)''$ are the two first derivatives
of $\log F$. Provided that $\log F$ is concave, which is the case
for probit and logit regressions, the Hessian of the log-likelihood
is clearly a negative definite matrix. Moreover, if we consider the
Gaussian prior, then the Hessian is of the log-posterior is also negative
(as the sum of two negative matrices, as Gaussian densities are log-concave).
We stick to the Gaussian prior for now. 

This suggests the following standard approach to compute the MAP (maximum
a posterior) estimator, that is the point $\betamap$ that maximises
the posterior density $p(\bbeta|\data)$: to use Newton-Raphson, that
is, to iterate 

\begin{equation}
\bbeta_{\mathrm{(new})}=\bbeta_{\mathrm{(old})}-\bm{H}^{-1}\left\{ \frac{\partial}{\partial\bbeta}\log p(\bbeta_{\mathrm{(old})}|\data)\right\} \label{eq:Newton-iter}
\end{equation}
until convergence is reached; here $\bm{H}$ is Hessian of the log
posterior at $\bbeta=\bbeta_{\mathrm{(old})}$, as computed above.
The iteration above corresponds to finding the zero of a local, quadratic
approximation of the log-posterior. Newton-Raphson typically works
very well (converges in a small number of iterations) when the function
to maximise is concave. A variant of this approach is 

We note two points in passing. First, one may obtain the MLE (maximum
likelihood estimator) by simply taking $p(\bbeta)=1$ above (i.e.
a Gaussian with infinite variance). But the MLE is not properly defined
when complete separation occurs, that is, there exists a hyperplane
that separates perfectly the two outcomes: $y_{i}\bbeta_{\mathrm{CS}}^{T}\bx_{i}\geq0$
for some $\bbeta_{\mathrm{CS}}$ and all $i\in1:N$. This remark gives
an extra incentive for performing Bayesian inference, or at least
MAP estimation, in cases where complete separation may occur, in particular
when the number of covariates is large \citep{firth1993bias,MR2655663}.

Variants of Newton-Raphson may be obtained by adapting automatically
the step size (e.g. update is $\bbeta_{\mathrm{(new})}=\bbeta_{\mathrm{(old})}-\lambda\bm{H}^{-1}\left\{ \frac{\partial}{\partial\bbeta}\log p(\bbeta_{\mathrm{(old})}|\data)\right\} $,
and step size $\lambda$ is determined by line search) or replacing
the Hessian $\bm{H}$ by some approximation. Some of these algorithms
such as IRLS (iterated reweighted least squares) have a nice statistical
interpretation. For our purposes however, these variants seem to show
roughly similar performance, so we will stick to the standard version
of Newton-Raphson.

\subsection{Posterior maximisation (Cauchy prior)\label{sub:Posterior-maximisation-Cauchy}}

The log-density of the Cauchy prior is not concave: 
\[
\log p(\bbeta)=-\sum_{j=1}^{p}\log\left(\pi\sigma_{j}\right)-\sum_{j=1}^{p}\log(1+\beta_{j}^{2}/\sigma_{j}^{2})
\]
for scales $\sigma_{j}$ chosen as explained in Section \ref{sub:Likelihood,-prior}.
Hence, the corresponding log-posterior is no longer guaranteed to
be concave, which in turn means that Newton-Raphson might fail to
converge.

However, we shall observe that, for most of the datasets considered
in this paper, Newton-Raphson does converge quickly even for our Cauchy
prior. In each case, we used as starting point for the Newton-Raphson
iterations the OLS (ordinary least square) estimate. We suspect what
happens is that, for most standard datasets, the posterior derived
from a Cauchy prior remains log-concave, at least in a region that
encloses the MAP estimator and our starting point.

\section{Fast approximation methods\label{sec:Fast-approximation-methods}}

This section discusses fast approximation methods, that is methods
that are deterministic, fast (compared to sampling-based methods),
but which comes with an approximation error which is difficult to
assess. These methods include the Laplace approximation, which was
popular in Statistics before the advent of MCMC methods, but also
recent Machine Learning methods, such as EP (Expectation Propagation,
\citealp{minka2001expectation}), and VB (Variational Bayes, e.g.
\citealp{Bishop:book}, Chap. 10). We will focus on Laplace and EP;
for VB, see \citet{consonni2007mean} for a discussion of why VB (or
at least a certain standard version of VB, known as mean field VB)
may not work so well for probit models.

Concretely, we will focus on the approximation of the following posterior
quantities: the marginal likelihood $p(\data)$, as this may be used
in model choice; and the marginal distributions $p(\beta_{i}|\data)$
for each component $\beta_{i}$ of $\bbeta.$ Clearly these are the
most commonly used summaries of the posterior distribution, and other
quantities, such as the posterior expectation of $\bbeta$, may be
directly deduced from them. 

Finally, one should bear in mind that such fast approximations may
be used as a preliminary step to calibrate an exact, more expensive
method, such as those described in Section \ref{sec:Exact-methods}.

\subsection{Laplace approximation\label{sub:Laplace-approximations}}

The Laplace approximation is based on a Taylor expansion of the posterior
log-density around the mode $\betamap$: 

\[
\log p(\bbeta|\data)\approx\log p(\betamap|\data)-\frac{1}{2}\left(\bbeta-\betamap\right)^{T}\bm{Q}\left(\bbeta-\betamap\right),
\]
where $\bm{Q}=-\bm{H}$, i.e. minus the Hessian of $\log p(\bbeta|\data)$
at $\bbeta=\betamap$; recall that we explained how to compute these
quantities in Section \ref{sub:Posterior-max-Gauss-prior}. One may
deduce a Gaussian approximation of the posterior by simply exponentiating
the equation above, and normalising: \texttt{
\begin{multline}
q_{L}(\bbeta)=N_{p}\left(\bbeta;\betamap,\bm{Q}^{-1}\right)\\
\eqdef(2\pi)^{-p/2}\left|\bm{Q}\right|^{1/2}\exp\left\{ -\frac{1}{2}\left(\bbeta-\betamap\right)^{T}\bm{Q}\left(\bbeta-\betamap\right)\right\} .\label{eq:LaplaceGaussApprox}
\end{multline}
}In addition, since for any $\bbeta$, 
\[
p(\data)=\frac{p(\bbeta)p(\data|\bbeta)}{p(\bbeta|\data)}
\]
one obtains an approximation to the marginal likelihood $p(\data)$
as follows: 
\[
p(\data)\approx Z_{L}(\data)\eqdef\frac{p(\betamap)p(\data|\betamap)}{(2\pi)^{-p/2}\left|\bm{Q}\right|^{1/2}}.
\]
From now on, we will refer to this particular Gaussian approximation
$q_{L}$ as the Laplace approximation, even if this phrase is sometimes
used in Statistics for higher-order approximations, as discussed in
the next Section. We defer to Section \ref{sub:Comparison-approxs}
the discussion of the advantages and drawbacks of this approximation
scheme.

\subsection{Improved Laplace, connection with INLA}

Consider the marginal distributions $p(\beta_{j}|\data)=\int p(\bbeta|\data)\dd\bbeta_{-j}$
for each component $\beta_{j}$ of $\bbeta$, where $\bbeta_{-j}$
is $\bbeta$ minus $\beta_{j}$. A first approximation may be obtained
by simply computing the marginals of the Laplace approximation $q_{L}$.
An improved (but more expensive) approximation may be obtained from:
\[
p(\beta_{j}|\data)\propto\frac{p(\bbeta)p(\data|\bbeta)}{p(\bbeta_{-j}|\beta_{j},\data)}
\]
which suggests to choose a fine grid of $\beta_{j}$ values (deduced
for instance from $q_{L}(\bbeta)$), and for each $\beta_{j}$ value,
compute a Laplace approximation of $p(\bbeta_{-j}|\beta_{j},\data)$,
by computing the mode $\hat{\bbeta}_{-j}(\beta_{j})$ and the Hessian
$\hat{\bm{H}}(\beta_{j})$ of $\log p(\bbeta_{-j}|\beta_{j},\data)$,
and then approximate (up to a constant) 
\[
p(\beta_{j}|\data)\approx q_{IL}(\beta_{j})\propto\frac{p\left(\hat{\bbeta}(\beta_{j})\right)p(\data|\hat{\bbeta}(\beta_{j}))}{\left|\hat{H}(\beta_{j})\right|^{1/2}}
\]
where $\hat{\bbeta}(\beta_{j})$ is the vector obtained by inserting
$\beta_{i}$ at position $i$ in $\hat{\bbeta}_{-j}(\beta_{j})$,
and IL stands for ``Improved Laplace''. One may also deduce posterior
expectations of functions of $\beta_{j}$ in this way. See also \citet{tierney1986accurate},
\citet{Kass} for higher order approximations for posterior expectations. 

We note in passing the connection to the INLA scheme of \citet{rue2009approximate}.
INLA applies to posteriors $p(\bm{\theta},\bm{x}|\data)$ where $\bm{x}$
is a latent variable such that $p(\bm{x}|\bm{\theta},\data)$ is close
to a Gaussian, and $\bm{\theta}$ is a low-dimensional hyper-parameter.
It constructs a grid of $\bm{\theta}-$values, and for each grid point
$\bm{\theta}_{j}$, it computes an improve Laplace approximation of
the marginals of $p(\bm{x}|\bm{\theta}_{j},\data)$. In our context,
$\bbeta$ may be identified to $\bm{x}$, $\bm{\theta}$ to an empty
set, and INLA reduces to the improved Laplace approximation described
above.

\subsection{The EM algorithm of \citet{MR2655663} (Cauchy prior)}

\citet{MR2655663} recommend against the Laplace approximation for
a Student prior (of which our Cauchy prior is a special case), because,
as explained in Section \ref{sub:Posterior-maximisation-Cauchy},
the corresponding log-posterior is not guaranteed to be concave, and
this might prevent Newton-Raphson to converge. In our simulations
however, we found the Laplace approximation to work reasonably well
for a Cauchy prior. We now briefly describe the alternative approximation
scheme proposed by \citet{MR2655663} for Student priors, which we
call for convenience Laplace-EM. 

Laplace-EM is based on the well-known representation of a Student
distribution, $\beta_{j}|\sigma_{j}^{2}\sim\Norm_{1}(0,\sigma_{j}^{2})$,
$\sigma_{j}^{2}\sim\InvGam(\nu/2,s_{j}\nu/2)$; take $\nu=1$ to recover
our Cauchy prior. Conditional on $\bsigsq=(\sigma_{1}^{2},\ldots,\sigma_{p}^{2})$,
the prior on $\bbeta$ is Gaussian, hence, for a fixed $\bsigsq$
one may implement Newton-Raphson to maximise the log-density of $p(\bbeta|\bsigsq,\data)$,
and deduce a Laplace (Gaussian) approximation of the same distribution.

Laplace-EM is an approximate EM \citep[Expectation Maximisation,][]{DemLaiRub}
algorithm, which aims at maximising in $\bsigsq=(\sigma_{1}^{2},\ldots,\sigma_{p}^{2})$
the marginal posterior distribution $p(\bsigsq|\data)=\int p(\bsigsq,\bbeta|\data)\,\dd\bbeta$.
Each iteration involves an expectation with respect to the intractable
conditional distribution $p(\bbeta|\bsigsq,\data)$, which is Laplace
approximated, using a single Newton-Raphson iteration. When this approximate
EM algorithm has converged to some value $\bsigsq_{\star}$, one more
Newton-Raphson iteration is performed to compute a final Laplace approximation
of $p(\bbeta|\bsigsq_{\star},\data)$, which is then reported as a
Gaussian approximation to the posterior. We refer the readers to \citet{MR2655663}
for more details on Laplace-EM.

\subsection{Expectation-Propagation\label{sub:Expectation-Propagation}}

Like Laplace, Expectation Propagation \citep[EP,][]{minka2001expectation}
generates a Gaussian approximation of the posterior, but it is based
on different ideas. The consensus in machine learning seems to be
that EP provides a better approximation than Laplace \citep[e.g. ][]{NickishRasmussen:ApproxGaussianProcClass};
the intuition being that Laplace is `too local' (i.e. it fitted so
at to match closely the posterior around the mode), while EP is able
to provide a global approximation to the posterior. 

Starting from the decomposition of the posterior as product of $(\ndata+1)$
factors:
\[
p(\bbeta|\data)=\frac{1}{p(\data)}\prod_{i=0}^{\ndata}l_{i}(\bbeta),\quad l_{i}(\bbeta)=F(y_{i}\bbeta^{T}\bx_{i})\mbox{ for }i\geq1,
\]
and $l_{0}$ is the prior, $l_{0}(\bbeta)=p(\bbeta)$, EP computes
iteratively a parametric approximation of the posterior with the same
structure 
\begin{equation}
q_{\mathrm{EP}}(\bbeta)=\prod_{i=0}^{\ndata}\frac{1}{Z_{i}}q_{i}(\bbeta).\label{eq:EP_dist}
\end{equation}
Taking $q_{i}$ to be an unnormalised Gaussian densities written in
natural exponential form 
\[
q_{i}(\bbeta)=\exp\left\{ -\frac{1}{2}\bbeta^{T}\bm{Q}_{i}\bbeta+\bbeta^{T}\bm{r}_{i}\right\} ,
\]
one obtains for $q_{EP}$ a Gaussian with natural parameters $\bm{Q}=\sum_{i=0}^{n}\bm{Q}_{i}$
and $\bm{r}_{i}=\sum_{i=0}^{n}\bm{r}_{i}$; note that the more standard
parametrisation of Gaussians may be recovered by taking 
\[
\bm{\Sigma}=\bm{Q}^{-1},\quad\bm{\mu}=\bm{Q}^{-1}\bm{r}.
\]
Other exponential families could be considered for $q$ and the $q_{i}$'s,
see e.g. \citet{Seeger:EPExpFam}, but Gaussian approximations seems
the most natural choice here. 

An EP iteration consists in updating one factor $q_{i}$, or equivalently
$\left(Z_{i},\bm{Q}_{i},\bm{r}_{i}\right)$, while keeping the other
factors as fixed, by moment matching between the hybrid distribution
\[
h(\bbeta)\propto l_{i}(\bbeta)\prod_{j\neq i}q_{j}(\bbeta)
\]
and the global approximation $q$ defined in (\ref{eq:EP_dist}):
compute 

\begin{eqnarray*}
Z_{h} & = & \int l_{i}(\bbeta)\prod_{j\neq i}q_{j}(\bbeta)\,\dd\bbeta\\
\bm{\mu}_{h} & = & \frac{1}{Z_{h}}\int\bbeta l_{i}(\bbeta)\prod_{j\neq i}q_{j}(\bbeta)\,\dd\bbeta\\
\bm{\Sigma}_{h} & = & \frac{1}{Z_{h}}\int\bbeta\bbeta^{T}l_{i}(\bbeta)\prod_{j\neq i}q_{j}(\bbeta)\,\dd\bbeta
\end{eqnarray*}
and set 
\[
\bm{Q}_{i}=\bm{\Sigma}_{h}^{-1}-\bm{Q}_{-i},\quad\bm{r}_{i}=\bm{\Sigma}_{h}^{-1}\bm{\mu}_{h}-\bm{r}_{-i},\quad\log Z_{i}=\log Z_{h}-\Psi(\bm{r},\bm{Q})+\Psi(\bm{r}_{-i},\bm{Q}_{-i})
\]
where $\bm{r}_{-i}=\sum_{j\neq i}\bm{r}_{j}$, $\bm{Q}_{-i}=\sum_{j\neq i}\bm{Q}_{j}$,
and $\psi(\bm{r},\bm{Q})$ is the normalising constant of a Gaussian
distribution with natural parameters $\left(\bm{r},\bm{Q}\right)$,
\[
\psi(\bm{r},\bm{Q})=\int_{\setR^{p}}\exp\left\{ -\frac{1}{2}\bbeta^{T}\bm{Q}\bbeta+\bbeta^{T}\bm{r}\right\} \,\dd\bbeta=-\frac{1}{2}\log\left|\bm{Q}/2\pi\right|+\frac{1}{2}\bm{r}^{T}\bm{Q}\bm{r}.
\]

In practice, EP proceeds by looping over sites, updating each one
in turn until convergence is achieved. 

To implement EP for binary regression models, two points must be addressed.
First, how to compute the hybrid moments? For the probit model, these
moments may be computed exactly, see the supplement, while for the
other links function (such as logistic), numerical (one-dimensional)
quadrature may be used. Second, how to deal with the prior? If the
prior is Gaussian, one may simply set $q_{0}$ to the prior, and never
update $q_{0}$ in the course of the algorithm. For a Cauchy prior,
$q_{0}$ is simply treated as an extra site. 

EP being a fairly recent method, it is currently lacking in terms
of supporting theory, both in terms of algorithmic convergence (does
it converge in a finite number of iterations?), and statistical convergence
(does the resulting approximation converges in some sense to the true
posterior distribution as $\ndata\rightarrow+\infty$?). On the other
hand, there is mounting evidence that EP works very well in many problems;
again see e.g. \citet[e.g. ][]{NickishRasmussen:ApproxGaussianProcClass}.

\subsection{Discussion of the different approximation schemes\label{sub:Comparison-approxs}}

Laplace and its variants have complexity $\OO(\ndata+p^{3})$, while
EP has complexity $\OO(\ndata p^{3})$. Incidentally, one sees that
the number of covariates $p$ is more critical than the number of
instances $\ndata$ in determining how `big' (how time-intensive to
process) is a given dataset. This will be a recurring point in this
paper. 

The $p^{3}$ term in both complexities is due to the $p\times p$
matrix operations performed by both algorithms; e.g. the Newton-Raphson
update (\ref{eq:Newton-iter}) requires solving a linear system of
order $p$. EP requires to perform such $p^{3}$ operations at each
site (i.e. for each single observation), hence the $\OO(\ndata p^{3})$
complexity, while Laplace perform such operations only once per iteration.
EP is therefore expected to be more expensive than Laplace. 

This remark may be mitigated as follows. First, one may modify EP
so as to update the global approximation only at the end of each iteration
(complete pass over the data). The resulting algorithm \citep{vanGerven2010}
may be easily implemented on parallel hardware: simply distribute
the $\ndata$ factors over the processors. Even without parallelisation,
parallel EP requires only one single matrix inversion per iteration. 

Second, the `improved Laplace' approximation for the marginals described
in Section \ref{sub:Laplace-approximations} requires to perform quite
a few basic Laplace approximations, so its speed advantage compared
to standard EP essentially vanishes. 

Points that remain in favour of Laplace is that it is simpler to implement
than EP, and the resulting code is very generic: adapting to either
a different prior, or a different link function (choice of $F$ in
\ref{eq:lik}), is simply a matter of writing a function that evaluates
the corresponding function. We have seen that such an adaptation requires
more work in EP, although to be fair the general structure of the
algorithm is not model-dependent. On the other hand, we shall see
that EP is often more accurate, and works in more examples, than Laplace;
this is especially the case for the Cauchy prior.

\section{Exact methods\label{sec:Exact-methods}}

We now turn to sampling-based methods, which are `exact', at least
in the limit: one may make the approximation error as small as desired,
by running the corresponding algorithm for long enough. We will see
that all of these algorithms requires some form of calibration that
requires prior knowledge on the shape of the posterior distribution.
Since the approximation methods covered in the previous section are
faster by orders of magnitude than sampling-based methods, we will
assume that a Gaussian approximation $q(\bbeta)$ (say, obtained by
Laplace or EP) has been computed in a preliminary step.

\subsection{Our gold standard: Importance sampling\label{sub:IS}}

Let $q(\bbeta)$ denote a generic approximation of the posterior $p(\bbeta|\data)$.
Importance sampling (IS) is based on the trivial identity 

\[
p(\data)=\int p(\bbeta)p(\data|\bbeta)\,\dd\bbeta=\int q(\bbeta)\frac{p(\bbeta)p(\data|\bbeta)}{q(\bbeta)}\,\dd\bbeta
\]
which leads to the following recipe: sample $\bbeta_{1},\ldots,\bbeta_{N}\sim q$,
then compute as an estimator of $p(\data)$ 
\begin{equation}
Z_{N}=\frac{1}{N}\sum_{n=1}^{N}w(\bbeta_{n}),\quad w(\bbeta)\eqdef\frac{p(\bbeta)p(\data|\bbeta)}{q(\bbeta)}.\label{eq:IS_evidence}
\end{equation}
In addition, since

\[
\int\varphi(\bbeta)p(\bbeta|\data)\,\dd\bbeta=\frac{\int\varphi(\bbeta)q(\bbeta)w(\bbeta)\,\dd\bbeta}{\int q(\bbeta)w(\bbeta)\,\dd\bbeta}
\]
one may approximate any posterior moment as 
\begin{equation}
\varphi_{N}=\frac{\sum_{n=1}^{N}w(\bbeta_{n})\varphi(\bbeta_{n})}{\sum_{n=1}^{N}w(\bbeta_{n})}.\label{eq:IS_postmoment}
\end{equation}
Approximating posterior marginals is also straightforward; one may
for instance use kernel density estimation on the weighted sample
$\left(\bbeta_{n},w(\bbeta_{n})\right)_{n=1}^{N}$. 

Concerning the choice of $q$, we will restrict ourselves to the Gaussian
approximations generated either from Laplace or EP algorithm. It is
sometimes recommended to use a Student distribution instead, as a
way to ensure that the variance of the above estimators is finite,
but we did not observe any benefit for doing so in our simulations. 

It is of course a bit provocative to call IS our gold standard, as
it is sometimes perceived as an obsolete method. We would like to
stress out however that IS is hard to beat relative to most of the
criteria laid out in the introduction: 
\begin{itemize}
\item because it is based on IID sampling, assessing the Monte Carlo error
of the above estimators is trivial: e.g. the variance of $Z_{N}$
may be estimated as $N^{-1}$ times the empirical variance of the
weights $w(\bbeta_{n})$. The auto-normalised estimator \ref{eq:IS_postmoment}
has asymptotic variance 
\[
\E_{q}\left[w(\bbeta)^{2}\left\{ \varphi(\bbeta)-\mu(\varphi)\right\} ^{2}\right],\quad\mu(\varphi)=\int\varphi(\bbeta)p(\bbeta|\data)\,\dd\bbeta
\]
which is also trivial to approximate from the simulated $\bbeta_{n}$'s.
\item Other advantages brought by IID sampling are: (a) importance sampling
is easy to parallelize; and (b) importance sampling is amenable to
QMC (Quasi-Monte Carlo) integration, as explained in the following
section. 
\item Importance sampling offers an approximation of the marginal likelihood
$p(\data$) at no extra cost. 
\item Code is simple and generic. 
\end{itemize}
Of course, what remains to determine is whether importance sampling
does well relative to our main criterion, i.e. error versus CPU trade-off.
We do know that IS suffers from a curse of dimensionality: take both
$q$ and and the target density $\pi$ to be the density of IID distributions:
$q(\bbeta)=\prod_{j=1}^{p}q_{1}(\beta_{j})$, $\pi(\bbeta)=\prod_{j=1}^{p}\pi_{1}(\beta_{j})$;
then it is easy to see that the variance of the weights grows exponentially
with $p$. Thus we expect IS to collapse when $p$ is too large; meaning
that a large proportion of the $\bbeta_{n}$ gets a negligible weight.
On the other hand, for small to moderate dimensions, we will observe
surprising good results; see Section \ref{sec:Numerics}. We will
also present below a SMC algorithm that automatically reduces to IS
when IS performs well, while doing something more elaborate in more
difficult scenarios. 

The standard way to assess the weight degeneracy is to compute the
effective sample size \citep{Kong1994}, 
\[
\mathrm{ESS}=\frac{\left\{ \sum_{n=1}^{N}w(\bbeta_{n})\right\} ^{2}}{\sum_{n=1}^{N}w(\bbeta_{n})^{2}}\in[1,N],
\]
which roughly approximates how many simulations from the target distribution
would be required to produce the same level of error. In our simulations,
we will compute instead the efficiency factor $\mathrm{EF}$, which
is simply the ratio $\mathrm{EF}=\mathrm{ESS}/N$.

\subsection{Improving importance sampling by Quasi-Monte Carlo\label{sub:QMC}}

Quasi-Monte Carlo may be seen as an elaborate variance reduction technique:
starting from the Monte Carlo estimators $Z_{N}$ and $\varphi_{N}$,
see (\ref{eq:IS_evidence}) and (\ref{eq:IS_postmoment}), one may
re-express the simulated vectors as functions of uniform variates
$\bu_{n}$ in $[0,1]^{d}$; for instance: 

\[
\bbeta_{n}=\bm{\mu}+\bm{C}\bm{\zeta}_{n},\quad\bm{\zeta}_{n}=\bm{\Phi^{-1}}(\bu_{n})
\]
where $\bm{\Phi^{-1}}$ is $\Phi^{-1}$, the $N(0,1)$ inverse CDF,
applied component-wise. Then, one replaces the $N$ vectors $\bu_{n}$
by a low-discrepancy sequence; that is a sequence of $N$ vectors
that spread more evenly over $[0,1]^{d}$; e.g. a Halton or a Sobol'
sequence. Under appropriate conditions, QMC error converges at rate
$\OO(N^{-1+\epsilon})$, for any $\epsilon>0$, to be compared with
the standard Monte Carlo rate $\OO_{P}(N^{-1/2})$. We refer to \citet{Lemieux:MCandQMCSampling}
for more background on QMC, as well as how to construct QMC sequences.

Oddly enough, the possibility to use QMC in conjunction with importance
sampling is very rarely mentioned in the literature; see however \citet{hormann2005quasi}.
More generally, QMC seems often overlooked in Statistics. We shall
see however that this simple IS-QMC strategy often performs very well. 

One drawback of IS-QMC is that we lose the ability to evaluate the
approximation error in a simple manner. A partial remedy is to use
randomised Quasi-Monte Carlo (RQMC), that is, the $\bu_{n}$ are generated
in such a way that (a) with probability one, $\bu_{1:N}$ is a QMC
point set; and (b) each vector $\bu_{n}$ is marginally sampled from
$[0,1]^{d}$. Then QMC estimators that are empirical averages, such
as $Z_{N}=N^{-1}\sum_{n=1}^{N}w(\bbeta_{n})$ become unbiased estimators,
and their error may be assessed through the empirical variance over
repeated runs. Technically, estimators that are ratios of QMC averages,
such as $\varphi_{N}$, are not unbiased, but for all practical purposes
their bias is small enough that assessing error through empirical
variances over repeated runs remains a reasonable approach.

\subsection{MCMC\label{sub:MCMC}}

The general principle of MCMC (Markov chain Monte Carlo) is to simulate
a Markov chain that leaves invariant the posterior distribution $p(\bbeta|\data)$;
see \citet{RobCas} for a general overview. Often mentioned drawbacks
of MCMC simulation are (a) the difficulty to parallelize such algorithms
(although see e.g. \citealp{Jacob2011} for an attempt at this problem);
(b) the need to specify a good starting point for the chain (or alternatively
to determine the burn-in period, that is, the length of the initial
part of the chain that should be discarded) and (c) the difficulty
to assess the convergence of the chain (that is, to determine if the
distribution of $\bbeta_{t}$ at iteration $t$ is sufficiently close
to the invariant distribution $p(\bbeta|\data)$). 

To be fair, these problems are not so critical for binary regression
models. Regarding (b), one may simply start the chain from the posterior
mode, or from a draw of one of the Gaussian approximations covered
in the previous section. Regarding (c) for most standard datasets,
MCMC converges reasonably fast, and convergence is easy to assess
visually. The main issue in practice is that MCMC generates correlated
random variables, and these correlations inflate the Monte Carlo variance.

\subsubsection{Gibbs sampling\label{sub:Gibbs}}

Consider the following data-augmentation formulation of binary regression:
\begin{eqnarray*}
z_{i} & = & \bm{\beta}^{T}\bm{x}_{i}+\epsilon_{i}\\
y_{i} & = & \sgn(z_{i})
\end{eqnarray*}
where $\bm{z}=(z_{1},\ldots,z_{\ndata})^{T}$ is a vector of latent
variables, and assume for a start that $\epsilon_{i}\sim N(0,1)$
(probit regression). One recognises $p(\bbeta|\bm{z},\data)$ as the
posterior of a linear regression model, which is tractable (for an
appropriate prior). This suggests to sample from $p(\bbeta,\bm{z}|\data)$
using Gibbs sampling \citep{Chib}: i.e. iterate the two following
steps: (a) sample from $\bm{z}|\bbeta,\data$; and (b) sample from
$\bbeta|\bm{z},\data$. 

For (a), the $z_{i}$'s are conditionally independent, and follows
a truncated Gaussian distribution 
\[
\ensuremath{p(z_{i}}|\bbeta,\data)\propto\Norm_{1}\left(z_{i};\bbeta^{T}\bx_{i},1\right)\ind\left\{ z_{i}y_{i}>0\right\} 
\]
which is easy to sample from \citep{chopin2011fast}. For Step (b)
and a Gaussian prior $\Norm_{p}(\bm{0},\bm{\Sigma}_{\mathrm{prior}})$,
one has, thanks to standard conjugacy properties: 
\[
\bbeta|\bm{z},\data\sim\Norm_{p}\left(\bm{\mu}_{\mathrm{post}}(\bm{z}),\bm{\Sigma}_{\mathrm{post}}\right),\quad\bm{\Sigma}_{\mathrm{post}}^{-1}=\bm{\Sigma}_{\mathrm{prior}}^{-1}+\bx\bx^{T},\quad\bm{\mu}_{\mathrm{post}}(\bm{z})=\bm{\Sigma}_{\mathrm{post}}^{-1}\bm{x}\bm{z}
\]
where $\bm{x}$ is the $n\times p$ matrix obtained by stacking the
$\bm{x}_{i}^{T}$. Note that $\bm{\Sigma}_{\mathrm{post}}$ and its
inverse need to be computed only once, hence the complexity of a Gibbs
iteration is $\OO(p^{2})$, not $\OO(p^{3})$. 

The main drawback of Gibbs sampling is that it is particularly not
generic: its implementation depends very strongly on the prior and
the model. Sticking to the probit case, switching to another prior
requires deriving a new way to update $\bbeta|\bm{z},\data$. For
instance, for a prior which is a product of Students with scales $\sigma_{j}$
(e.g. our Cauchy prior), one may add extra latent variables, by resorting
to the well-known representation: $\beta_{j}|s_{j}\sim\Norm_{1}(0,\nu\sigma_{j}^{2}/s_{j})$,
$s_{j}\sim\Chisq(\nu)$; with $\nu=1$ for our Cauchy prior. Then
the algorithm has three steps: (a) an update of the $z_{i}$'s, exactly
as above; (b) an update of $\bbeta$, as above but with $\bm{\Sigma}_{\mathrm{prior}}$
replaced by the diagonal matrix with elements $\nu\sigma_{j}^{2}/s_{j}$,
$j=1,\ldots,p$; and (c) an (independent) update of the $p$ latent
variables $s_{j}$, with $s_{j}|\bbeta,\bz,\data\sim\Gam\left((1+\nu)/2,\left(1+\nu\beta_{j}^{2}/\sigma_{j}^{2}\right)/2\right)$.
The complexity of Step (b) is now $\OO(p^{3})$, since $\bm{\Sigma}_{\mathrm{prior}}$
and $\bm{\Sigma}_{\mathrm{post}}$ must be recomputed at each iteration
(although some speed-up may be obtained by using Sherman\textendash Morrison
formula). 

Of course, considering yet another type of prior would require deriving
another strategy for sampling $\bbeta$. Then if one turns to logistic
regression, things get rather complicated. In fact, deriving an efficient
Gibbs sampler for logistic regression is a topic of current research;
see \citet{Holmes2006,FrhwirthSchnatter2009,Gramacy2012,Polson2013}.
In a nutshell, the two first papers use the same data augmentation
as above, but with $\epsilon_{i}\sim\mathrm{Logistic}(1)$ written
as a certain mixture of Gaussians (infinite for the first paper, finite
but approximate for the second paper), while \citet{Polson2013} use
instead a representation of a logistic likelihood as an infinite mixture
of Gaussians, with a Polya-Gamma as the mixing distribution. Each
representation leads to introducing extra latent variables, and discussing
how to sample their conditional distributions. 

Since their implementation is so model-dependent, the main justification
for Gibbs samplers should be their greater performance relative to
more generic algorithms. We will investigate if this is indeed the
case in our numerical section.

\subsubsection{Hastings-Metropolis}

Hastings-Metropolis consists in iterating the step described as Algorithm
\ref{alg:Hastings-Metropolis-iteration}. Much like importance sampling,
Hastings-Metropolis is both simple and generic, that is, up to the
choice of the proposal kernel $\kappa(\bbeta^{\star}|\bbeta)$ (the
distribution of the proposed point $\bbeta^{\star}$, given the current
point $\bbeta$). A naive approach is to take $\kappa(\bbeta^{\star}|\bbeta)$
independent of $\bbeta$, $\kappa(\bbeta^{\star}|\bbeta)=q(\bbeta^{\star})$,
where $q$ is some approximation of the posterior. In practice, this
usually does not work better than importance sampling based on the
same proposal, hence this strategy is hardly used. 

\begin{algorithm}
\begin{description}
\item [{Input}] $\bbeta$
\item [{Output}] $\bbeta'$
\item [{1}] Sample $\bbeta^{\star}\sim\kappa(\bbeta^{\star}|\bbeta)$.
\item [{2}] With probability $1\wedge r$, 
\[
r=\frac{p(\bbeta^{\star})p(\data|\bbeta^{\star})\kappa(\bbeta|\bbeta^{\star})}{p(\bbeta)p(\data|\bbeta)\kappa(\bbeta^{\star}|\bbeta)},
\]
set $\bbeta'=\bbeta^{\star}$; otherwise set $\bbeta'=\bbeta$. 
\end{description}
\protect\caption{\label{alg:Hastings-Metropolis-iteration}Hastings-Metropolis iteration}
\end{algorithm}

A more usual strategy is to set the proposal kernel to a random walk:
$\kappa(\bbeta^{\star}|\bbeta)=\Norm_{p}(\bbeta,\bm{\Sigma}_{\mathrm{prop}})$.
It is well known that  the choice of $\bm{\Sigma}_{\mathrm{prop}}$
is critical for good performance. For instance, in the univariate
case, if $\bm{\Sigma}_{\mathrm{prop}}$ is too small, the chain moves
slowly, while if too large, proposed moves are rarely accepted. 

A result from the optimal scaling literature \citep[e.g. ][]{RobertsRosenthal:OptimalScalingMH}
is that, for a $\Norm_{p}(\bm{0},\bm{I}_{p})$ target, $\bm{\Sigma}_{\mathrm{prop}}=(\lambda^{2}/p)\bm{I}_{p}$
with $\lambda=2.38$ is asymptotically optimal, in the sense that
as $p\rightarrow\infty$, this choice leads to the fastest exploration.
Since the posterior of a binary regression model is reasonably close
to a Gaussian, we adapt this result by taking $\bm{\Sigma}_{\mathrm{prop}}=(\lambda^{2}/p)\bm{\Sigma}_{q}$
in our simulations, where $\bm{\Sigma}_{q}$ is the covariance matrix
of a (Laplace or EP) Gaussian approximation of the posterior. This
strategy seems validated by the fact we obtain acceptance rates close
to the optimal rate, as given by \citet{RobertsRosenthal:OptimalScalingMH}. 

The bad news behind this optimality result is that the chain requires
$\OO(p)$ steps to move a $\OO(1)$ distance. Thus random walk exploration
tends to become slow for large $p$. This is usually cited as the
main motivation to develop more elaborate MCMC strategies, such as
HMC, which we cover in the following section.

\subsubsection{HMC}

Hamiltonian Monte Carlo (HMC, also known as Hybrid Monte Carlo, \citealp{Duane1987})
is a new type of MCMC algorithm, where one is able to perform several
steps in the parameter space before determining if the new position
is accepted or not. Consequently, HMC is able to make much bigger
jumps in the parameter space than standard Metropolis algorithms.
See \citet{Neal2010HMC} for an excellent introduction. 

Consider the pair $(\bbeta,\mom)$, where $\bbeta\sim p(\bbeta|\data)$,
and $\mom\sim N_{p}(0,M^{-1})$, thus with joint un-normalised density
$\exp\left\{ -H(\bbeta,\mom)\right\} $, with

\[
H(\bbeta,\mom)=E(\bbeta)+\frac{1}{2}\mom^{T}\bm{M}\mom,\quad E(\bbeta)=-\log\left\{ p(\bbeta)p(\data|\bbeta)\right\} .
\]
The physical interpretation of HMC is that of a particle at position
$\bbeta$, with velocity $\mom$, potential energy $E(\bbeta)$, kinetic
energy $ $$\frac{1}{2}\mom^{T}M\mom$, for some mass matrix $M$,
and therefore total energy given by $H(\bbeta,\mom)$. The particle
is expected to follow a trajectory such that $H(\bbeta,\mom)$ remains
constant over time. 

In practice, HMC proceeds as follows: first, sample a new velocity
vector, $\mom\sim N_{p}(0,M^{-1})$. Second, move the particle while
keeping the Hamiltonian $H$ constant; in practice, discretisation
must be used, so $L$ steps of step-size $\epsilon$ are performed
through leap-frop steps; see Algorithm \ref{alg:Leapfrog} which describes
one such step. Third, the new position, obtained after $L$ leap-frog
steps is accepted or rejected according to probability $1\wedge\exp\left\{ H(\bbeta,\mom)-H(\bbeta^{\star},\mom^{\star})\right\} $;
see Algorithm \ref{alg:HMC-iteration} for a summary. The validity
of the algorithm relies on the fact that a leap-frog step is ``volume
preserving''; that is, the deterministic transformation $\left(\bbeta,\mom\right)\rightarrow\left(\bbeta_{1},\mom_{1}\right)$
has Jacobian one. This is why the acceptance probability admits this
simple expression. 

\begin{algorithm}
\begin{description}
\item [{Input}] $\left(\bbeta,\mom\right)$
\item [{Output}] $\left(\bbeta_{1},\mom_{1}\right)$
\item [{1}] $\mom_{1/2}\leftarrow\mom-\frac{\epsilon}{2}\nabla_{\bbeta}E(\bbeta)$
\item [{2}] $\bbeta_{1}\leftarrow\bbeta+\epsilon\mom_{1/2}$
\item [{3}] $\mom_{1}\leftarrow\mom_{1/2}-\frac{\epsilon}{2}\nabla_{\bbeta}E(\bbeta_{1})$
\end{description}
\protect\caption{\label{alg:Leapfrog}Leap-frog step}
\end{algorithm}

\begin{algorithm}
\begin{description}
\item [{Input}] $\bbeta$
\item [{Output}] $\bbeta'$
\item [{1}] Sample momentum $\mom\sim\Norm_{p}(0,\bm{M}).$
\item [{2}] Perform $L$ leap-frog steps (see Algorithm \ref{alg:Leapfrog}),
starting from $\left(\bbeta,\mom\right)$; call $\left(\bbeta^{\star},\mom^{\star}\right)$
the final position. 
\item [{3}] With probability $1\wedge r$, 
\[
r=\exp\left\{ H(\bbeta,\mom)-H(\bbeta^{\star},\mom^{\star})\right\} 
\]
set $\bbeta^{'}=\bbeta^{\star}$; otherwise set $\bbeta^{'}=\bbeta$. 
\end{description}
\protect\caption{\label{alg:HMC-iteration}HMC iteration}
\end{algorithm}

The tuning parameters of HMC are $\bm{M}$ (the mass matrix), $L$
(number of leap-frog steps), and $\epsilon$ (the stepsize). For $\bm{M}$,
we follow \citet{Neal2010HMC}'s recommendation and take $\bm{M}^{-1}=\bm{\Sigma}_{q}$,
an approximation of the posterior variance (again obtained from either
Laplace or EP). This is equivalent to rescaling the posterior so as
to have a covariance matrix close to identity. In this way, we avoid
the bad mixing typically incurred by strong correlations between components. 

The difficulty to choose $L$ and $\epsilon$ seems to be the main
drawback of HMC. The performance of HMC seems very sensitive to these
tuning parameters, yet clear guidelines on how to choose them seem
currently lacking. A popular approach is to fix $L\epsilon$ to some
value, and to use vanishing adaptation \citep{andrieu2008tutorial}
to adapt $\epsilon$ so as to target acceptance rate of $0.65$ (the
optimal rate according to the formal study of HMC by \citealp{Beskos2013}):
i.e. at iteration $t$, take $\epsilon=\epsilon_{t}$, with $\epsilon_{t}=\epsilon_{t-1}-\eta_{t}(R_{t}-0.65)$,
$\eta_{t}=t^{-\kappa}$, $\kappa\in(1/2,1)$ and $R_{t}$ the acceptance
rate up to iteration $t$. The rationale for fixing $L\epsilon$ is
that quantity may be interpreted as a `simulation length', i.e. how
much distance one moves at each step; if too small, the algorithm
may exhibit random walk behaviour, while if too large, it may move
a long distance before coming back close to its starting point. Since
the spread of is already taken into account through $\bm{M}^{-1}=\bm{\Sigma}_{q}$,
we took $\epsilon L=1$ in our simulations.

\subsubsection{NUTS and other variants of HMC}

\citet{girolami2011riemann} proposed an interesting variation of
HMC, where the mass matrix $\bm{M}$ is allowed to depends on $\bbeta$;
e.g. $\bm{M}(\bbeta)$ is set to the Fisher information of the model.
This allows the corresponding algorithm, called RHMC (Riemanian HMC),
to adapt locally to the geometry of the target distribution. The main
drawback of RHMC is that each iteration involves computing derivatives
of $M(\bbeta)$ with respect to $\bbeta$, which is very expensive,
especially if $p$ is large. For binary regression, we found RMHC
to be too expensive relative to plain HMC, even when taking into account
the better exploration brought by RHMC. This might be related to the
fact that the posterior of a binary regression model is rather Gaussian-like
and thus may not require such a local adaptation of the sampler. 

We now focus on NUTS \citep[No U-Turn sampler,][]{NUTS}, a variant
of HMC which does not require to specify a priori $L$, the number
of leap-frog steps. Instead, NUTS aims at keeping on doing such steps
until the trajectory starts to loop back to its initial position.
Of course, the difficulty in this exercise is to preserve the time
reversibility of the simulated Markov chain. To that effect, NUTS
constructs iteratively a binary tree whose leaves correspond to different
velocity-position pairs $(\bm{\alpha},\bbeta)$ obtained after a certain
number of leap-frog steps. The tree starts with two leaves, one at
the current velocity-position pair, and another leaf that corresponds
to one leap-frop step, either in the forward or backward direction
(i.e. by reversing the sign of velocity); then it iteratively doubles
the number of leaves, by taking twice more leap frog steps, again
either in the forward or backward direction. The tree stops growing
when at least one leaf corresponds to a ``U-turn''; then NUTS chooses
randomly one leaf, among those leaves that would have generated the
current position with the same binary tree mechanism; in this way
reversibility is preserved. Finally NUTS moves the new position that
corresponds to the chosen leaf. 

We refer the readers to \citet{NUTS} for a more precise description
of NUTS. Given its complexity, implementing directly NUTS seems to
require more efforts than the other algorithms covered in this paper.
Fortunately, the STAN package (\url{http://mc-stan.org/}) provides
a C++ implementation of NUTS which is both efficient and user-friendly:
the only required input is a description of the model in a probabilistic
programming language similar to BUGS. In particular, STAN is able
to automatically derive the log-likelihood and its gradient, and no
tuning of any sort is required from the user. Thus, we will use STAN
to assess NUTS in our numerical comparisons.

\subsection{Sequential Monte Carlo\label{sub:Sequential-Monte-Carlo}}

Sequential Monte Carlo (SMC) is a class of algorithms for approximating
iteratively a sequence of distributions $\pi_{t}$, $t=0,\ldots,T$,
using importance sampling, resampling, and MCMC steps. We focus here
on the non-sequential use of SMC \citep{Neal:AIS,Chopin:IBIS,DelDouJas:SMC},
where one is only interested in approximating the final distribution
$\pi_{T}$ (in our case, set to the posterior $p(\bbeta|\data)$),
and the previous $\pi_{t}$'s are designed so as to allow for a smooth
progression from some $\pi_{0}$, which is easy to sample from, to
$\pi_{T}$.

At iteration $t$, SMC produces a set of weighted particles (simulations)
$(\bbeta_{n},w_{n})_{n=1}^{N}$ that approximates $\pi_{t}$, in the
sense that 
\[
\frac{1}{\sum_{n=1}^{N}w_{n}}\sum_{n=1}^{N}w_{n}\varphi(\bbeta_{n})\rightarrow\E^{\pi_{t}}\left[\varphi(\bbeta)\right]
\]
as $N\rightarrow+\infty$. At time $0$, one samples $\bbeta^{n}\sim\pi_{0}$,
and set $w_{n}=1$. To progress from $\pi_{t-1}$ to $\pi_{t}$, one
uses importance sampling: weights are multiplied by ratio $\pi_{t}(\bbeta_{n})/\pi_{t-1}(\bbeta_{n})$.
When the variance of the weights gets too large (which indicates that
too few particles contribute significantly to the current approximation),
one resamples the particles: each particle gets reproduced $O_{n}$
times, where $O_{n}\geq0$ is random, and such that $\E(O_{n})=Nw_{n}/\sum_{m=1}^{N}w_{m}$,
and $\sum_{n=1}^{N}O_{n}=N$ with probability one. In this way, particles
with a low weights are likely to die, while particles with a large
weight get reproduced many times. Finally, one may re-introduce diversity
among the particles by applying one (or several) MCMC steps, using
a MCMC kernel that leaves invariant the current distribution $\pi_{t}$. 

We focus in this paper on tempering SMC, where the sequence 
\[
\pi_{t}(\bbeta)\propto q(\bbeta)^{1-\delta_{t}}\left\{ p(\bbeta)p(\data|\bbeta)\right\} ^{\delta_{t}}
\]
corresponds to a linear interpolation (on the log-scale) between some
distribution $\pi_{0}=q$, and $\pi_{T}(\bbeta)=p(\bbeta|\data)$,
our posterior. This is a convenient choice in our case, as we have
at our disposal some good approximation $q$ (either from Laplace
or EP) of our posterior. A second advantage of tempering SMC is that
one can automatically adapt the ``temperature ladder'' $\delta_{t}$
\citep{jasrainference}. Algorithm \ref{alg:tempering-SMC} describes
a tempering SMC algorithm based on such an adaptation scheme: at each
iteration, the next distribution $\pi_{t}$ is chosen so that the
efficiency factor (defined in Section \ref{sub:IS}) of the importance
sampling step from $\pi_{t-1}$ to $\pi_{t}$ equals a pre-defined
level $\tau\in(0,1)$; a default value is $\tau=1/2$. 

\begin{algorithm}
\begin{raggedright}
Operations involving index $n$ must be performed for all $n\in1:N$. 
\par\end{raggedright}
\begin{description}
\item [{0}] Sample $\bbeta_{n}\sim q(\bbeta)$ and set $\underbar{\ensuremath{\delta}}\leftarrow0$.
\item [{1}] Let, for $\delta\in[\underbar{\ensuremath{\delta}},1]$, 
\[
\mathrm{EF}(\delta)=\frac{1}{N}\frac{\left\{ \sum_{n=1}^{N}w_{\gamma}(\bbeta_{n})\right\} ^{2}}{\left\{ \sum_{n=1}^{N}w_{\gamma}(\bbeta_{n})^{2}\right\} },\quad u_{\delta}(\bbeta)=\left\{ \frac{p(\bbeta)p(\data|\bbeta)}{q(\bbeta)}\right\} ^{\delta}.
\]
If $\mathrm{\,EF}(1)\geq\tau$, stop and return $(\bbeta_{n},w_{n})_{n=1:N}$
with $w_{n}=u_{1}(\bbeta_{n})$; otherwise, use the bisection method
\citep[Chap. 9]{NumericalRecipes3rdEd} to solve numerically in $\delta$
the equation $\mathrm{EF}(\gamma)=\tau.$
\item [{2}] Resample according to normalised weights $W_{n}=w_{n}/\sum_{m=1}^{N}w_{m}$,
with $w_{n}=u_{\delta}(\bbeta_{n})$; see the supplement for one such
resampling algorithm.
\item [{3}] Update the $\bbeta_{n}$'s through $m$ MCMC steps that leaves
invariant $\pi_{t}(\bbeta)$, using e.g. Algorithm \ref{alg:Hastings-Metropolis-iteration}
with $\kappa(\bbeta^{\star}|\bbeta)=\Norm_{p}(\bbeta,\bm{\Sigma}_{\mathrm{prop}})$,
$\bm{\Sigma}_{\mathrm{prop}}=\lambda\hat{\bm{\Sigma}}$, where $\hat{\bm{\Sigma}}$
is the empirical covariance matrix of the resampled particles. 
\item [{4}] Set $\underbar{\ensuremath{\delta}}\leftarrow\delta$. Go to
Step 1. 
\end{description}
\protect\caption{\label{alg:tempering-SMC}tempering SMC}
\end{algorithm}

Another part of Algorithm \ref{alg:tempering-SMC} which is easily
amenable to automatic calibration is the MCMC step. We use a random
walk Metropolis step, i.e. Algorithm \ref{alg:Hastings-Metropolis-iteration}
with proposal kernel $\kappa(\bbeta^{\star}|\bbeta)=\Norm_{p}(\bbeta,\bm{\Sigma}_{\mathrm{prop}})$,
 but with $\bm{\Sigma}_{\mathrm{prop}}$ calibrated to the empirical
variance of the particles $\hat{\bm{\Sigma}}$: $\bm{\Sigma}_{\mathrm{prop}}=\lambda\hat{\bm{\Sigma}}$,
for some $\lambda$. Finally, one may also automatically calibrate
the number $m$ of MCMC steps, as in \citet{ridgway2014computation},
but in our simulations we simply took $m=3$. 

In the end, one obtains essentially a black-box algorithm. In practice,
we shall often observe that, for simple datasets, our SMC algorithm
automatically reduces to a single importance sampling step, because
the efficiency factor of moving from the initial distribution $q$
to the posterior is high enough. In that case, our SMC sampler performs
exactly as standard importance sampling.

Finally, we note that the reweighting step and the MCMC steps of Algorithm
\ref{alg:tempering-SMC} are easy to parallelise.

\section{Numerical study\label{sec:Numerics}}

The point of this section is to compare numerically the different
methods discussed in the previous sections, first on several datasets
of standard size (that are representative of previous numerical studies),
then in a second time on several bigger datasets. 

We focus on the following quantities: the marginal likelihood of the
data, $p(\data)$, and the $p$ marginal posterior distributions of
the regression coefficients $\beta_{j}$. Regarding the latter, we
follow \citet{Faes2011} in defining the `marginal accuracy' of approximation
$q$ for component $j$ to be 
\[
\mathrm{MA}_{j}=1-\frac{1}{2}\int_{-\infty}^{+\infty}\left|q(\beta_{j})-p(\beta_{j}|\data)\right|\,\dd\beta_{j}.
\]
This quantity lies in $[0,1]$, and is scale-invariant. Since the
true marginals $p(\beta_{j}|\data)$ are not available, we will approximate
them through a Gibbs sampler run for a very long time. To give some
scale to this criterion, assume $q(\beta_{j})=\Norm_{1}(\beta_{j};\mu_{1},\sigma^{2})$,
$p(\beta_{j}|\data)=\Norm_{1}(\beta_{j};\mu_{2},\sigma^{2})$, then
$\mathrm{MA}_{j}$ is $2\Phi(-\delta/2)\approx1-0.4\times\delta$
for $\delta=|\mu_{1}-\mu_{2}|/\sigma$ small enough; e.g. $0.996$
for $\delta\approx0.01$, $0.96$ for $\delta\approx0.1$. 

In our results, we will refer to the following four prior/model `scenarios':
Gaussian/probit, Gaussian/logit, Cauchy/probit, Cauchy/logit, where
Gaussian and Cauchy refer to the two priors discussed in Section \ref{sub:Likelihood,-prior}.
All the algorithms have been implemented in C++, using the Armadillo
and Boost libraries, and run on a standard desktop computer (except
when explicitly stated). Results for NUTS were obtained by running
STAN (\url{http://mc-stan.org/}) version 2.4.0.

\subsection{Datasets of moderate size}

Table \ref{tab:Datasets-of-moderate-size} lists the 7 datasets considered
in this section (obtained from the UCI machine learning repository,
except Elections, which is available on the web page of \citet{gelman2006data}'s
book). These datasets are representative of the numerical studies
found in the literature. In fact, it is a super-set of the real datasets
considered in \citet{girolami2011riemann}, \citet{shahbaba2011split},
\citet{Holmes2006} and also (up to one dataset with 5 covariates)
\citet{Polson2013}. In each case, an intercept have been included;
i.e. $p$ is the number of predictors plus one. 

\begin{table}
\begin{centering}
\begin{tabular}{lcc}
\hline 
Dataset & $\ndata$ & $p$\tabularnewline
\hline 
Pima (Indian diabetes) & 532 & 8\tabularnewline
German (credit) & 999 & 25\tabularnewline
Heart (Statlog) & 270 & 14\tabularnewline
Breast (cancer) & 683 & 10\tabularnewline
Liver (Indian Liver patient) & 579 & 11\tabularnewline
Plasma (blood screening data) & 32 & 3\tabularnewline
Australian (credit)  & 690 & 15\tabularnewline
Elections & 2015 & 52\tabularnewline
\hline 
\end{tabular}
\par\end{centering}

\protect\caption{\label{tab:Datasets-of-moderate-size}Datasets of moderate size (from
UCI repository, except Elections, from web-site of \citet{gelman2006data}'s
book): name (short and long version), number of instances $\ndata$,
number of covariates $p$ (including an intercept)}
\end{table}

\subsubsection{Fast Approximations}

We compare the four approximation schemes described in Section \ref{sec:Fast-approximation-methods}:
Laplace, Improved Laplace, Laplace EM, and EP. We concentrate on the
Cauchy/logit scenario for two reasons: (i) Laplace EM requires a Student
prior; and (ii) Cauchy/logit seems the most challenging scenario for
EP, as (a) a Cauchy prior is more difficult to deal with than a Gaussian
prior in EP ; and (b) contrary to the probit case, the site update
requires some approximation; see Section \ref{sub:Expectation-Propagation}
for more details. 

\begin{figure}
\centering{}\includegraphics[scale=0.35]{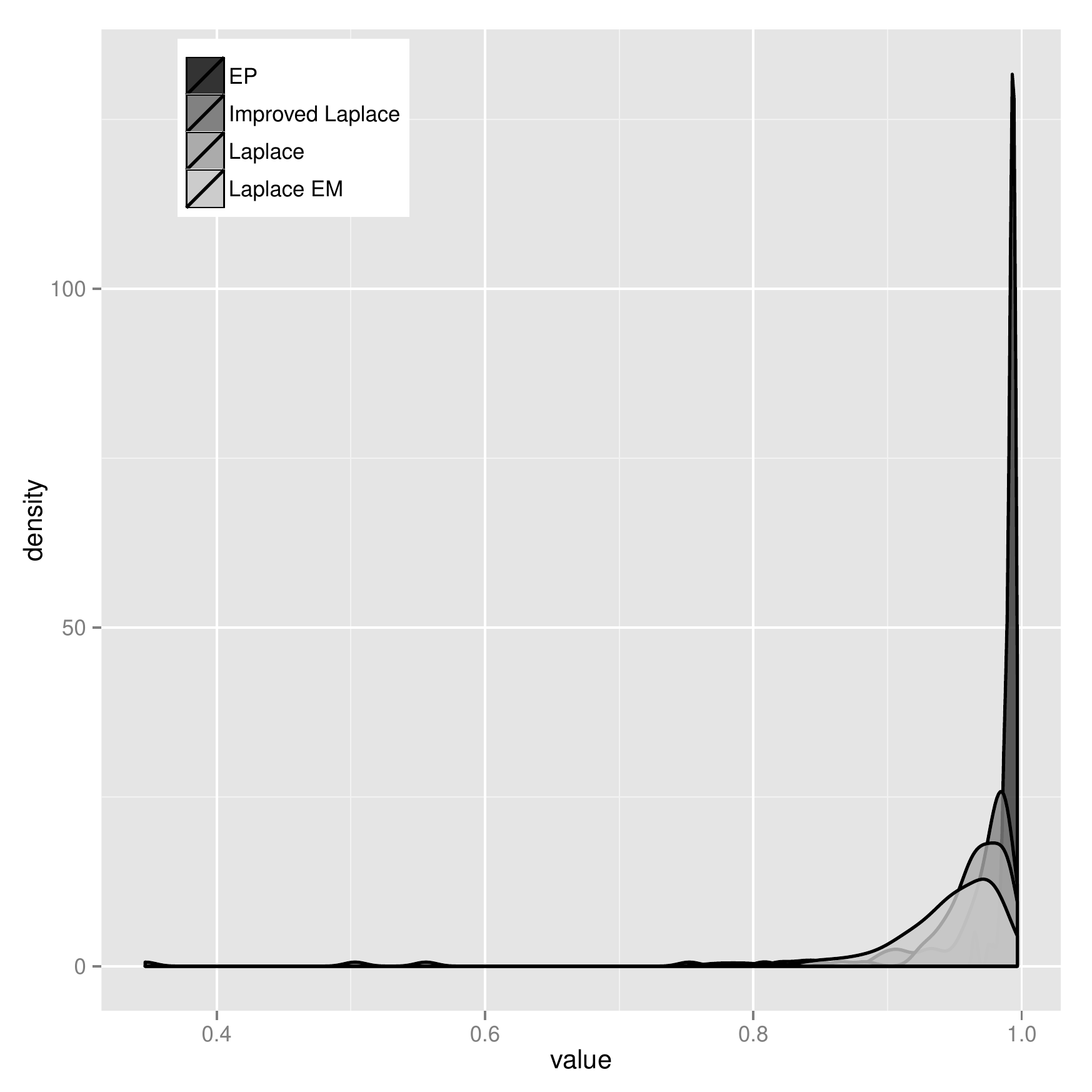}\includegraphics[scale=0.35]{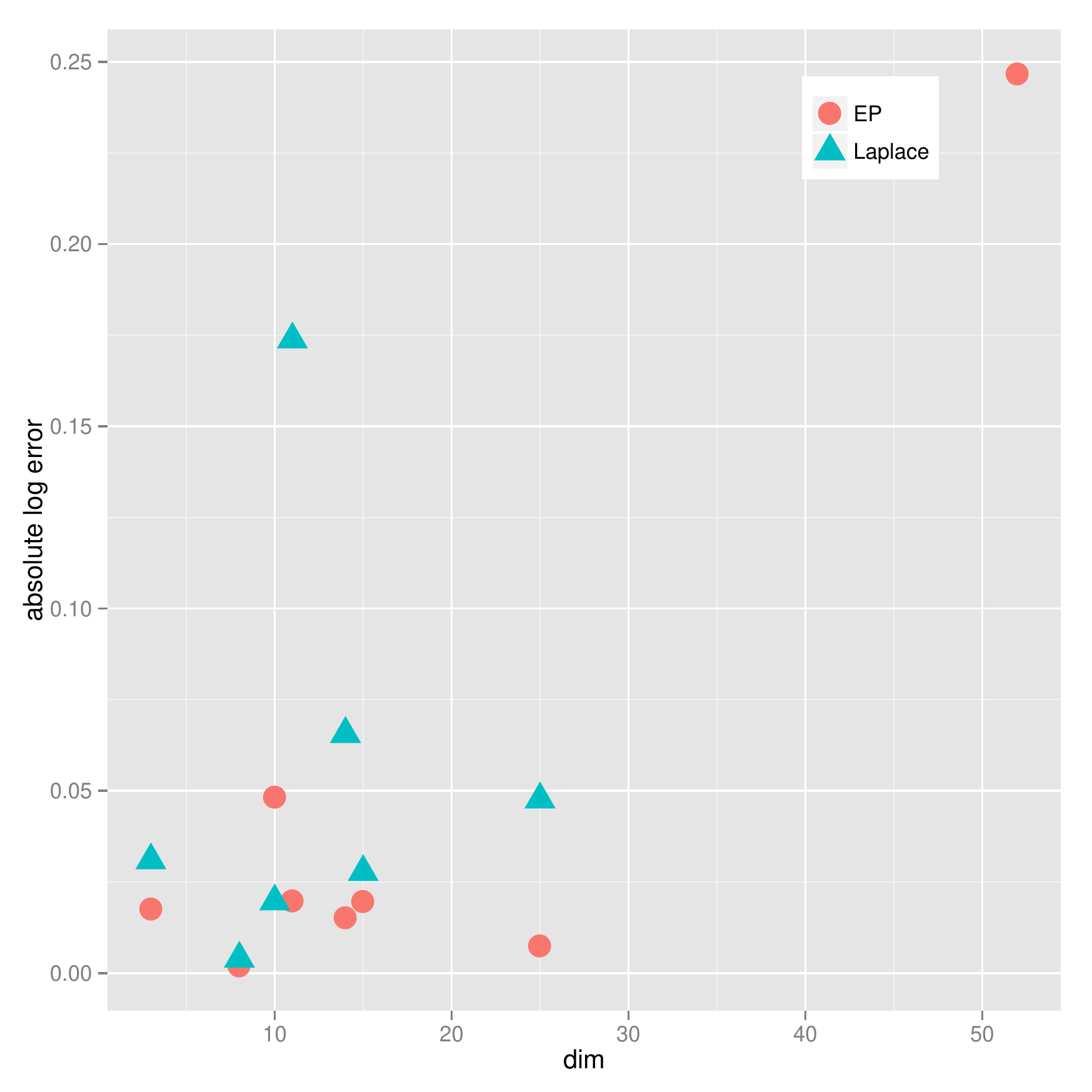}\protect\caption{\label{fig:approx-alldata}Comparison of approximation schemes across
all datasets of moderate size: marginal accuracies (left), and absolute
error for log-evidence versus the dimension $p$ (right); $x-$axis
range of the left plot determined by range of marginal accuracies
(i.e. marginal accuracy may drop below $0.4$ for e.g. Laplace-EM).}
\end{figure}

\begin{figure}
\subfloat[Pima]{

\protect\centering{}\protect\includegraphics[scale=0.35]{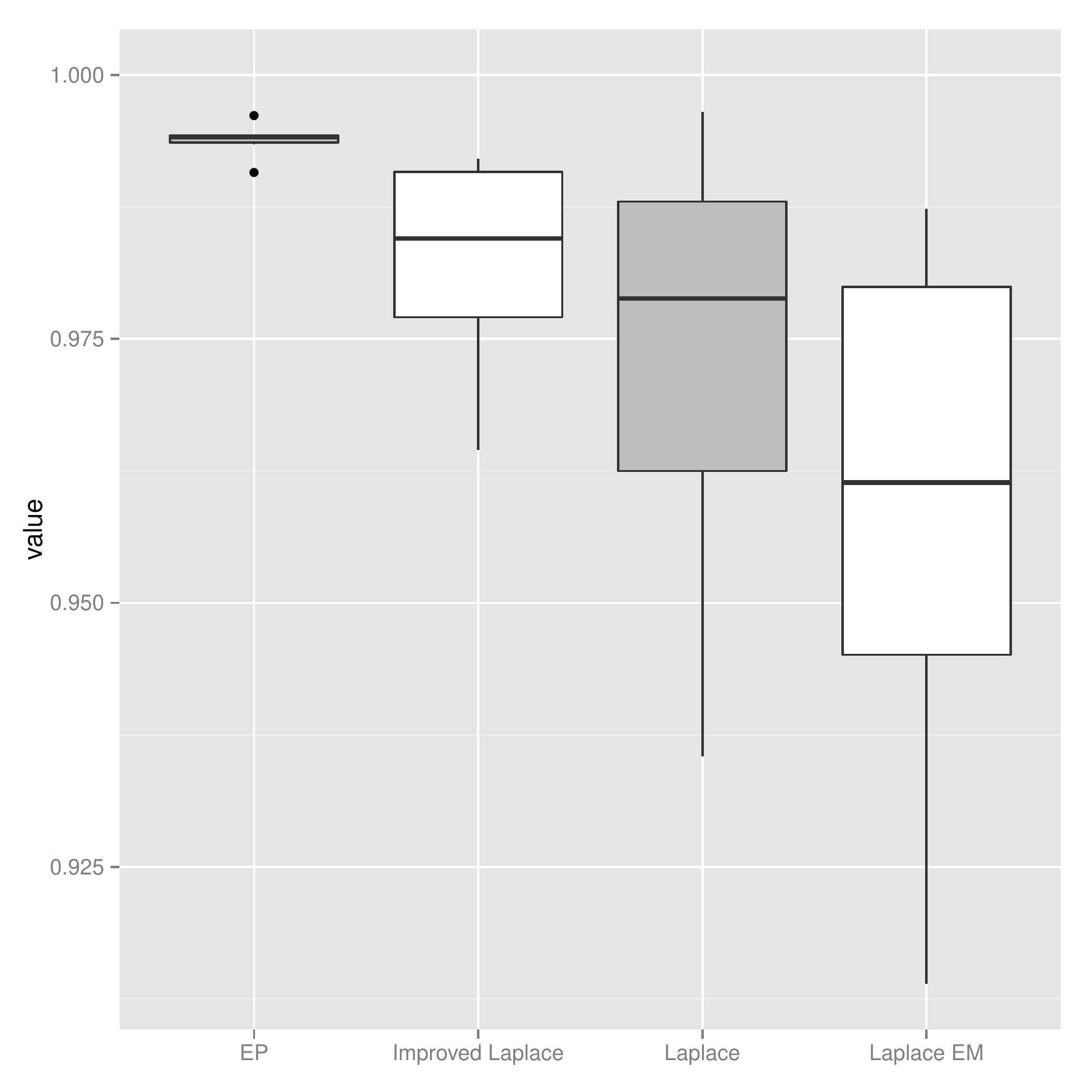}\protect}\subfloat[Heart]{

\protect\centering{}\protect\includegraphics[scale=0.35]{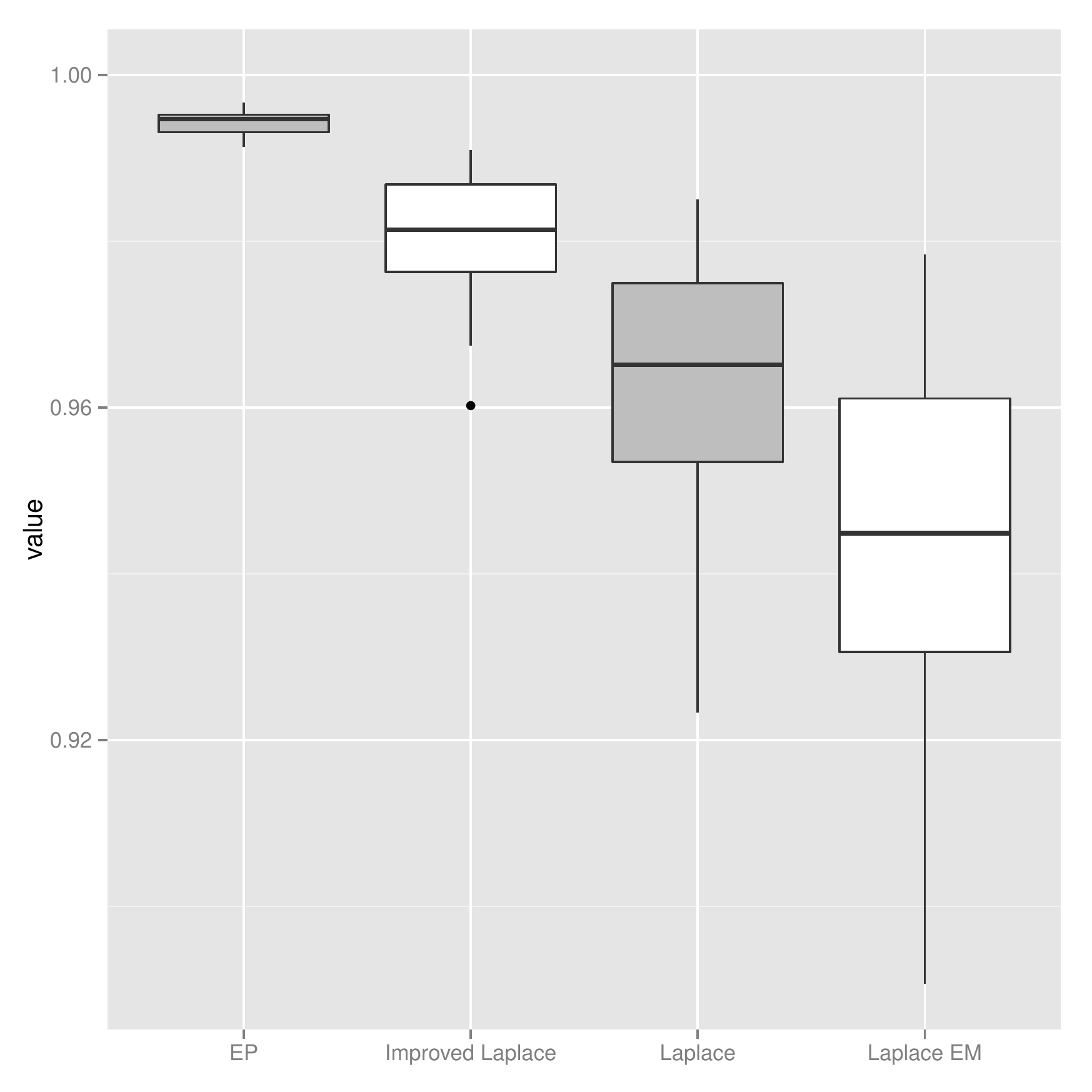}\protect}

\subfloat[Breast]{

\protect\centering{}\protect\includegraphics[scale=0.35]{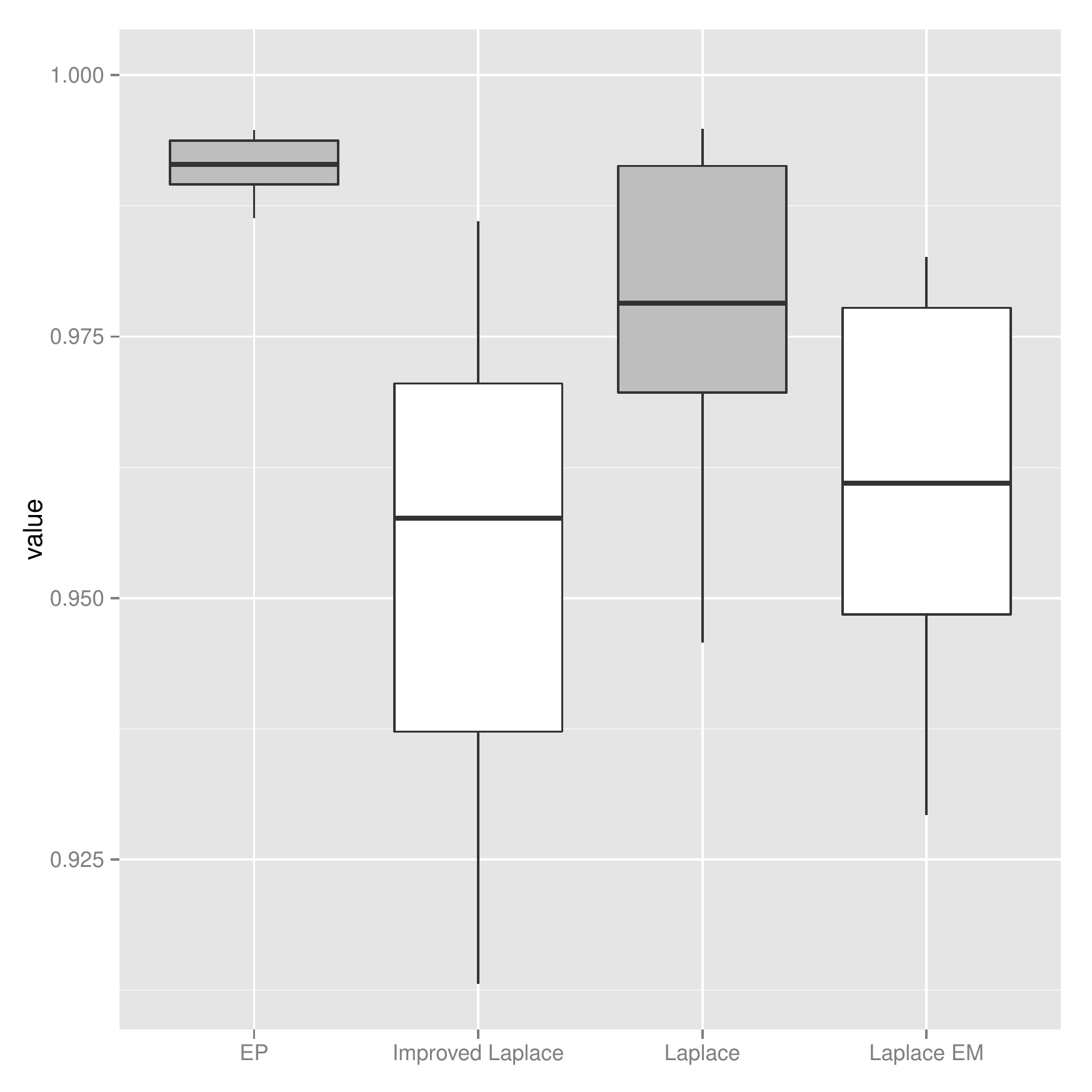}\protect}\subfloat[German]{

\protect\centering{}\protect\includegraphics[scale=0.35]{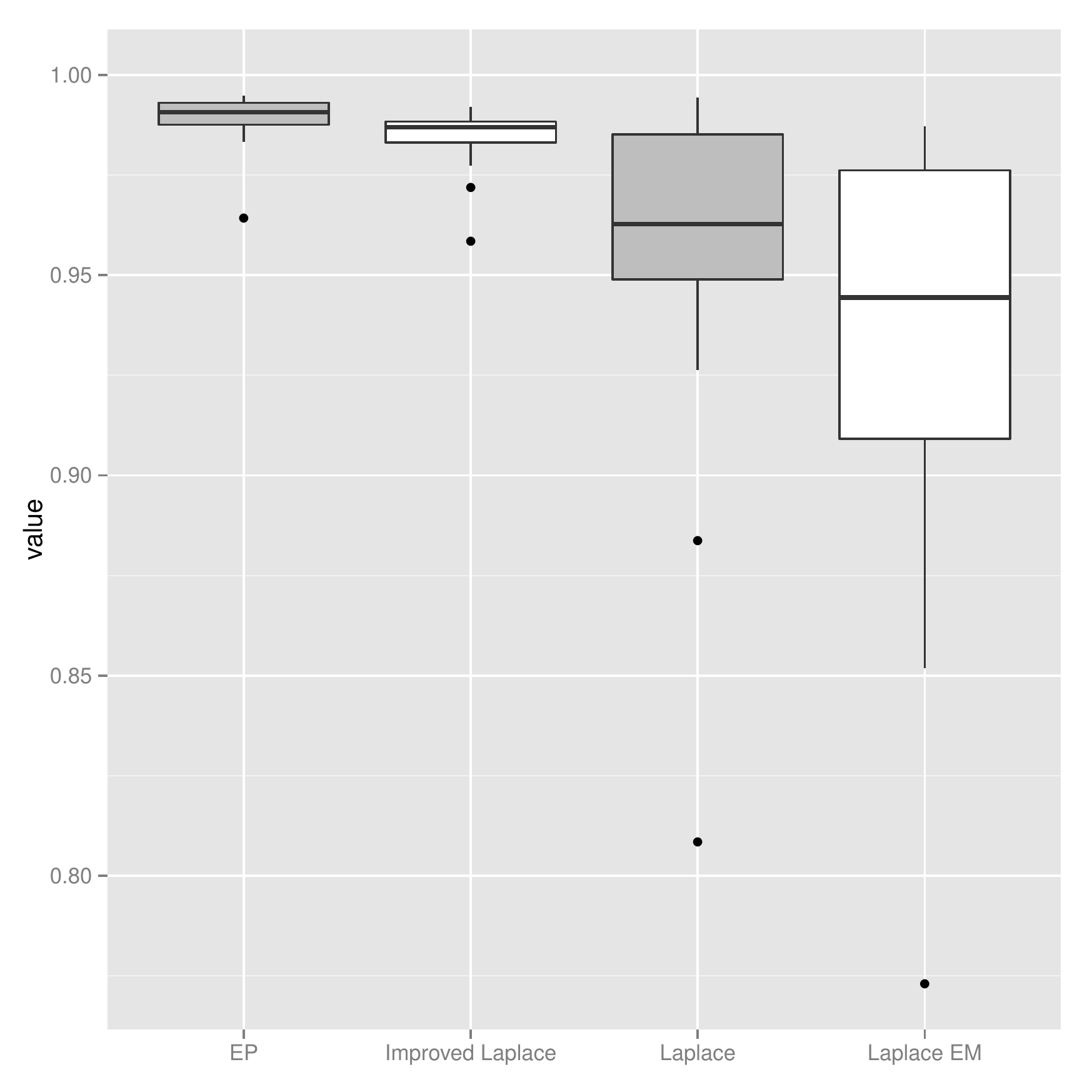}\protect}

\protect\caption{\label{fig:approx-selected-data}Box-plots of marginal accuracies
across the $p$ dimensions, for the four approximation schemes, and
four selected datasets; plots for remaining datasets are in the supplement.
For the sake of readability, scale of $y-$axis varies across plots. }
\end{figure}

Left panel of Fig. \ref{fig:approx-alldata} plots the marginal accuracies
of the four approximation schemes across all components and all datasets;
Fig. \ref{fig:approx-selected-data} does the same, but separately
for four selected datasets; results for the remaining datasets are
available in the supplement. 

EP seems to be the most accurate method on these datasets: marginal
accuracy is about $0.99$ across all components for EP, while marginal
accuracy of the other approximation schemes tend to be lower, and
may even drop to quite small values; see e.g. the German dataset,
and the left tail in the left panel of Fig. \ref{fig:approx-alldata}. 

EP also fared well in terms of CPU time: it was at most seven times
as intensive as standard Laplace across the considered datasets, and
about $10$ to $20$ times faster than Improved Laplace and Laplace
EM. As expected (see Section \ref{sub:Comparison-approxs}). Of course,
the usual caveats apply regarding CPU time comparison, and how they
may depend on the hardware, the implementation, and so on.

We also note in passing the disappointing performance of Laplace EM,
which was supposed to replace standard Laplace when the prior is Student,
but which actually performs not as well as standard Laplace on these
datasets. 

We refer the reader to the supplement for similar results on the three
other scenarios, which are consistent with those above. In addition,
we also represent the approximation error of EP and Laplace for approximating
the log-evidence in the right panel of Fig. \ref{fig:approx-alldata}.
Again, EP is found to be more accurate than Laplace for most datasets
(except for the Breast dataset). 

To conclude, it seems that EP may be safely be used as a complete
replacement of sampling-based methods on such datasets, as it produces
nearly instant results, and the approximation error along all dimensions
is essentially negligible.

\subsubsection{Importance sampling, QMC}

We now turn to importance sampling (IS), which we deemed our ``gold
standard'' among sampling-based methods, because of its ease of use
and other nice properties as discussed in Section \ref{sub:IS}. We
use $N=5\times10^{5}$ samples, and a Gaussian EP proposal. (Results
with a Laplace proposal are roughly similar.) We consider first the
Gaussian/probit scenario, because this is particularly favorable to
Gibbs sampling; see next section. Table \ref{tab:IS} reports for
each dataset the efficiency factor of IS (as defined in Section \ref{sub:IS}),
the CPU time and two other quantities discussed below.

\begin{table}[h]
\begin{centering}
\begin{tabular}{l|ccc|cc}
\hline 
 & \multicolumn{3}{c|}{IS} & \multicolumn{2}{c}{IS-QMC}\tabularnewline
\hline 
Dataset  & EF  & CPU  & MT  & MSE improv. & MSE improv.\tabularnewline
 & $=\mathrm{ESS}/N$ & time & speed-up & (expectation) & (evidence)\tabularnewline
\hline 
Pima  & 99.5\%  & 37.54 s  & 4.39  & 28.9  & 42.7 \tabularnewline
German  & 97.9\%  & 79.65 s  & 4.51  & 13.2  & 8.2\tabularnewline
Breast  & 82.9\%  & 50.91 s  & 4.45  & 2.6  & 6.2 \tabularnewline
Heart  & 95.2\%  & 22.34 s  & 4.53 & 8.8  & 9.3 \tabularnewline
Liver  & 74.2 \%  & 35.93 s  & 4.76  & 7.6  & 11.3 \tabularnewline
Plasma  & 90.0\%  & 2.32 s  & 4.28  & 2.2  & 4.4\tabularnewline
Australian  & 95.6\%  & 53.32 s  & 4.57  & 12  & 20.3\tabularnewline
Elections & 21.39\% & 139.48 s & 3.87 & 617.9 & 3.53\tabularnewline
\hline 
\end{tabular}
\par\end{centering}

\protect\caption{\label{tab:IS} Performance of importance sampling (IS), and QMC importance
sampling (IS-QMC), on all datasets, in Gaussian/probit scenario: efficiency
factor (EF), CPU time (in seconds), speed gain when using multi-threading
Intel hyper-threaded quad core CPU (Speed gain MT), and efficiency
gain of QMC (see text). }
\end{table}

We see that all these efficiency factors are all close to one, which
means IS works almost as well as IID sampling would on such datasets.
Further improvement may be obtained by using either parallelization,
or QMC (Quasi-Monte Carlo, see Section \ref{sub:QMC}). Table \ref{tab:IS}
reports the speed-up factor obtained when implementing multi-threading
on our desktop computer which has a multi threading quad core CPU
(hence 8 virtual cores). We also implemented IS on an Amazon EC2 instance
with 32 virtual CPUs, and obtained speed-up factors about 20, and
running times below $2s$.

Finally, Table \ref{tab:IS} also reports the MSE improvement (i.e.
MSE ratio of IS relative to IS-QMC) obtained by using QMC, or more
precisely RQMC (randomised QMC), based on a scrambled Sobol' sequence
\citep[see e.g. ][]{Lemieux:MCandQMCSampling}. Specifically, the
table reports the median MSE improvement for the $p$ posterior expectations
(first column), and the MSE improvement for the evidence (second column).
The improvement brought by RQMC varies strongly across datasets.

The efficiency gains brought by parallelization and QMC may be combined,
because the bulk of the computation (as reported by a profiler) is
the $N$ likelihood evaluations, which are trivial to parallelize. 

It is already clear that other sampling-based methods do not really
have a fighting chance on such datasets, but we shall compare them
in the next section for the sake of completeness. See also the supplement
for results for other scenarios, which are very much in line with
those above.

\subsubsection{MCMC schemes}

\begin{figure}
\begin{centering}
\subfloat[{Median IRIS for the $p$ posterior expectations $\mathbb{E}[\beta_{j}|\data]$ }]{\protect\begin{centering}
\protect\includegraphics[scale=0.35]{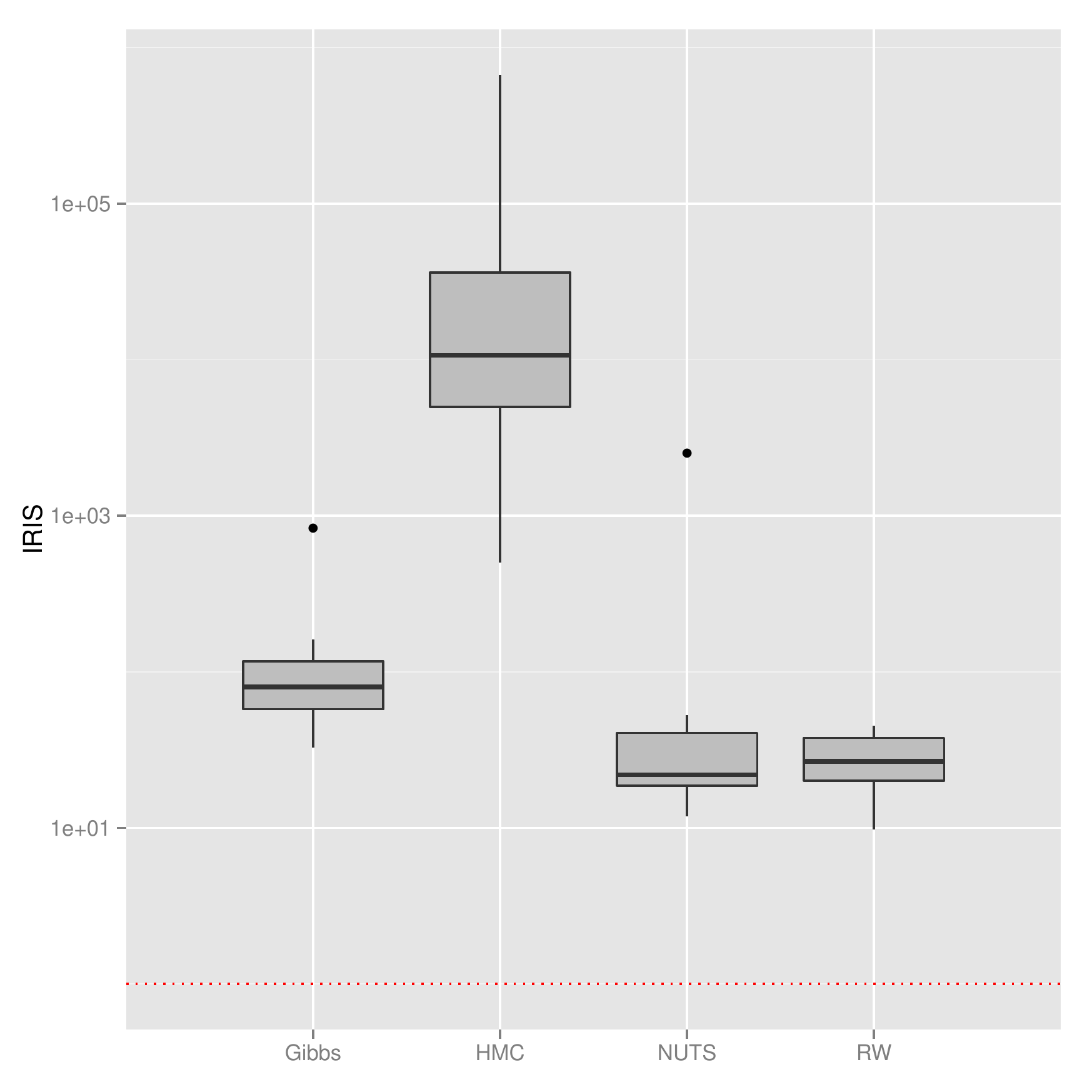}\protect
\par\end{centering}

}\subfloat[{Median IRIS for the $p$ posterior variances $\mathrm{Var}[\beta_{j}|\data]$}]{\protect\begin{centering}
\protect\includegraphics[scale=0.35]{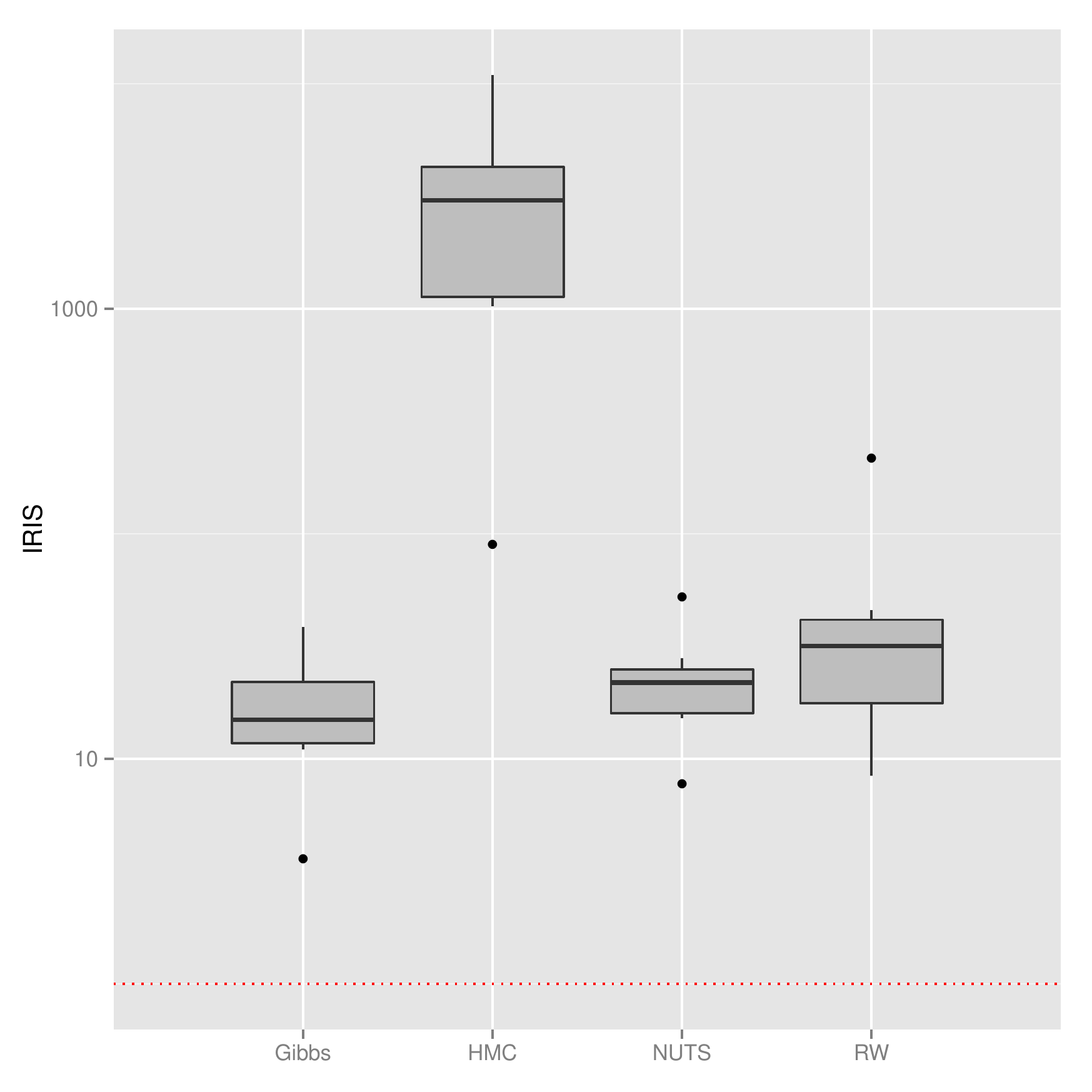}\protect
\par\end{centering}

}
\par\end{centering}

\protect\caption{\label{fig:IRIS-MCMC}IRIS (Inefficiency relative to importance sampling)
across all datasets for MCMC schemes and Gaussian/probit scenario;
left (resp. right) panel shows median IRIS when estimating the $p$
posterior expectations (resp. the $p$ posterior variances). }
\end{figure}

In order to compare the different sampling-based methods, we define
the IRIS (Inefficiency Relative to Importance Sampling) criterion,
for a given method $M$ and a given posterior estimate, as follows:
\[
\frac{\mathrm{MSE}_{M}}{MSE_{IS}}\times\frac{\mathrm{CPU}_{IS}}{\mathrm{CPU}_{M}}
\]
where $\mathrm{MSE}_{M}$ (resp. $\mathrm{MSE}_{IS}$) is the mean
square error of the posterior estimate obtained from method M (resp.
from importance sampling), and $\mathrm{CPU}_{M}$ the CPU time of
method M (resp. importance sampling). The comparison is relative to
importance sampling \emph{without} parallelisation or quasi-Monte
Carlo sampling. In terms of posterior estimates, we consider the expectation
and variance of each posterior marginal $p(\beta_{j}|\data)$. We
observe that, in both cases, IRIS does not vary much across the $p$
components, so we simply report the median of these $p$ values. Fig
\ref{fig:IRIS-MCMC} reports the median IRIS across all datasets.
We refer the reader to Section \ref{sub:MCMC} for how we tuned these
MCMC algorithms. 

The first observation is that all these MCMC schemes are significantly
less efficient than importance sampling on such datasets. The source
of inefficiency seems mostly due to the autocorrelations of the simulated
chains (for Gibbs or random walk Metropolis), or, equivalently, the
number of leap-frog steps performed at each iteration in HMC and NUTS.
See the supplement for ACF's (Autocorrelation plots) to support this
statement. 

Second, HMC and NUTS do not perform significantly better than random-walk
Metropolis. As already discussed, HMC-type algorithms are expected
to outperform random walk algorithms as $p\rightarrow+\infty$. But
the considered datasets seem too small to give evidence to this phenomenon,
and should not be considered as reasonable benchmarks for HMC-type
algorithms (not to mention again that these algorithms are significantly
outperformed by IS on such datasets). We note in passing that it might
be possible to get better performance for HMC by finely tuning the
quantities $\epsilon$ and $L$ on per dataset basis. We have already
explained in the introduction why we think this is bad practice, and
we also add at this stage that the fact HMC requires so much more
effort to obtain good performance (relative to other MCMC samplers)
is a clear drawback. 

Regarding Gibbs sampling, it seems a bit astonishing that an algorithm
specialised to probit regression is not able to perform better than
more generic approach on such simple datasets. Recall that the Gaussian/probit
case is particularly favourable to Gibbs, as explained in Section
\ref{sub:Gibbs}. See the supplement for a comparison of MCMC schemes
in other scenarios than Gaussian/probit; results are roughly similar,
except that Gibbs is more significantly outperformed by other methods,
as expected.

\subsection{Bigger datasets}

Finally, we turn our attention to the bigger datasets summarised by
Table \ref{tab:larger-datasets}. These datasets not only have more
covariates (than those of the previous section), but also stronger
correlations between these covariates (especially Sonar and Musk).
We consider the probit/Gaussian scenario. 

\begin{table}
\begin{centering}
\begin{tabular}{lcc}
\hline 
Dataset & $\ndata$ & $p$\tabularnewline
\hline 
Musk & 476 & 95\tabularnewline
Sonar & 208 & 61\tabularnewline
DNA & 400 & 180\tabularnewline
\hline 
\end{tabular}
\par\end{centering}

\protect\caption{\label{tab:larger-datasets}Datasets of larger size (from UCI repository):
name, number of instances $\ndata$, number of covariates $p$ (including
an intercept)}
\end{table}

\begin{figure}
\subfloat[Musk]{\protect\begin{centering}
\protect\includegraphics[scale=0.35]{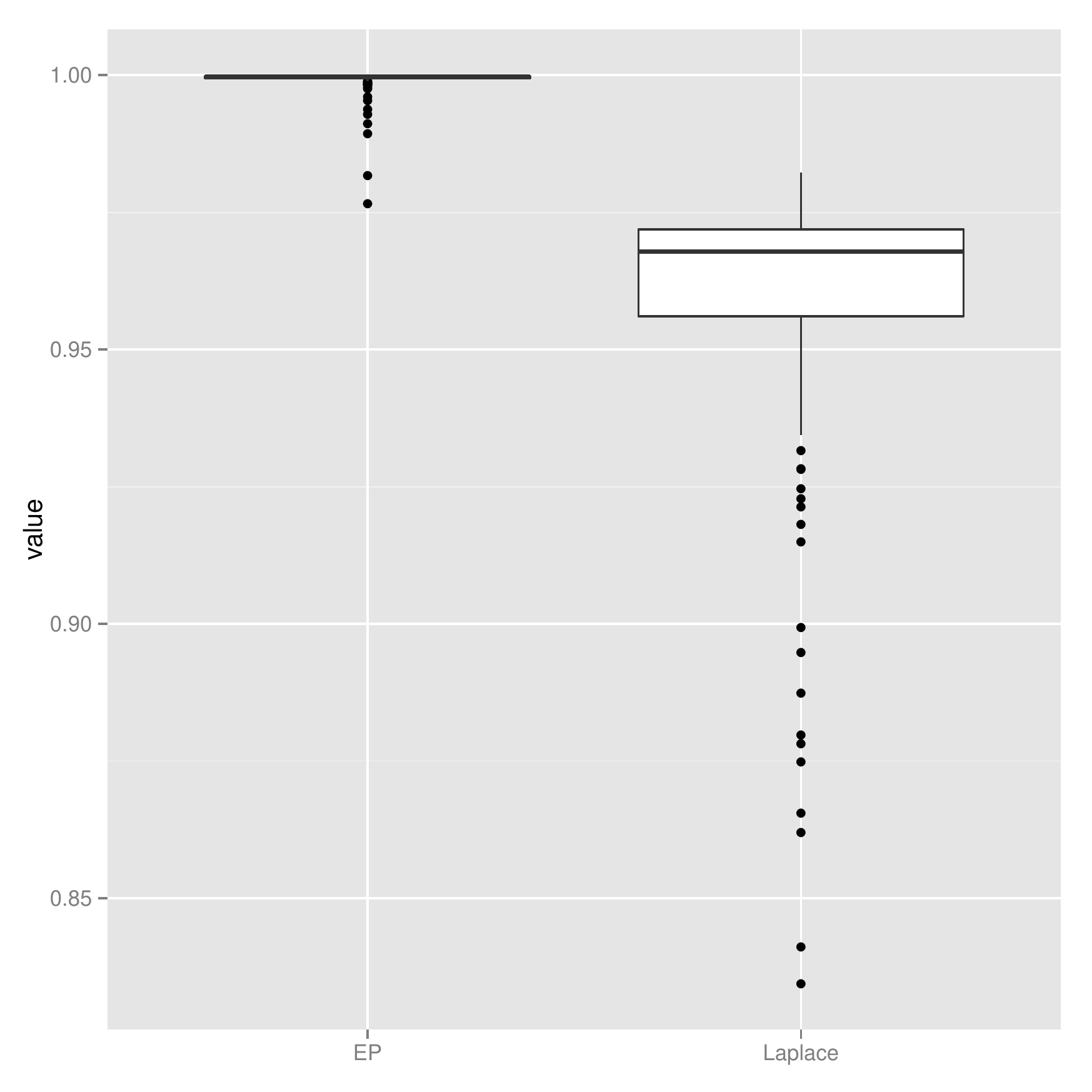}\protect
\par\end{centering}

}\subfloat[Sonar]{\protect\centering{}\protect\includegraphics[scale=0.35]{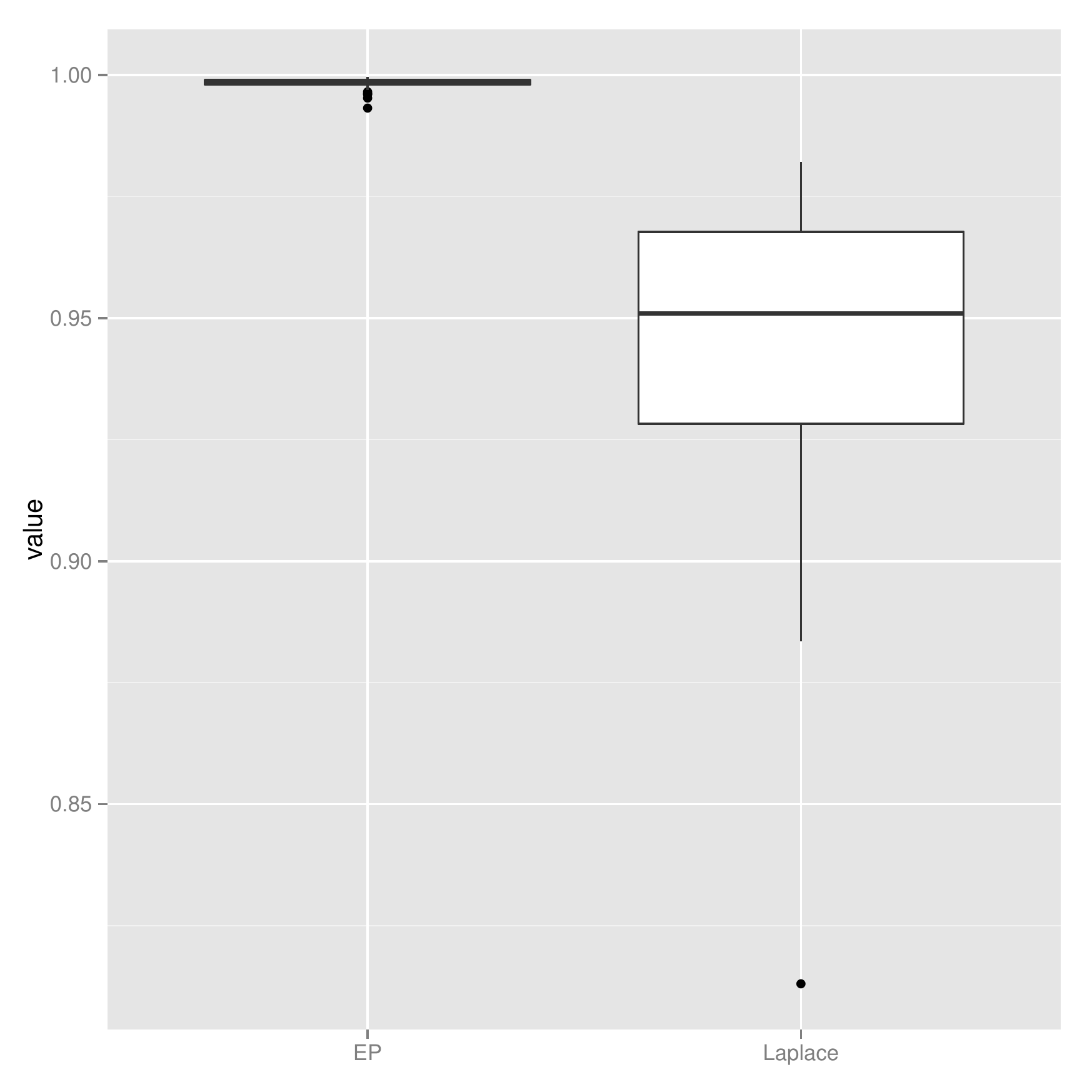}\protect}

\subfloat[DNA]{\protect\centering{}\protect\includegraphics[scale=0.35]{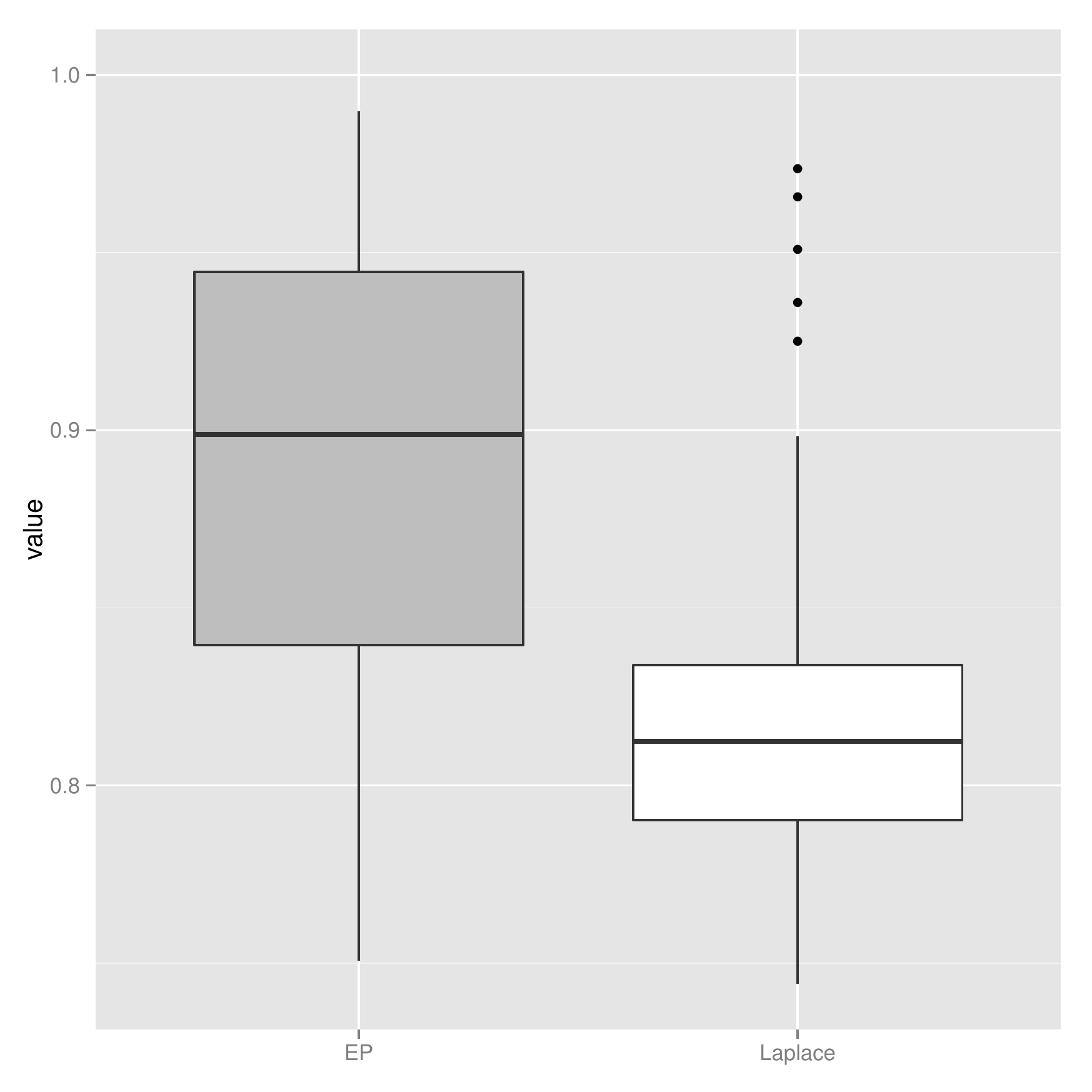}\protect}

\protect\caption{\label{fig:approx_big}Marginal accuracies across the $p$ dimensions
of EP and Laplace, for datasets Musk, Sonar and DNA}

\end{figure}

Regarding fast approximations, we observe again that EP performs very
well, and better than Laplace; see Figure \ref{fig:approx_big}. It
is only for DNA (180 covariates) that the EP approximation starts
to suffer. 

Regarding sampling-based methods, importance sampling may no longer
be used as a reference, as the effective sample size collapses to
a very small value for these datasets. We replace it by the tempering
SMC algorithm described in Section \ref{sub:Sequential-Monte-Carlo}.
Moreover, we did not manage to calibrate HMC so as to obtain reasonable
performance in this setting. Thus, among sampling-based algorithms,
the four remaining contenders are: Gibbs sampling, NUTS, RWHM (random
walk Hastings-Metropolis), and tempering SMC. Recall that the last
two are calibrated with the approximation provided by EP. 

\begin{figure}

\subfloat[Musk]{\protect\begin{centering}
\protect\includegraphics[scale=0.35]{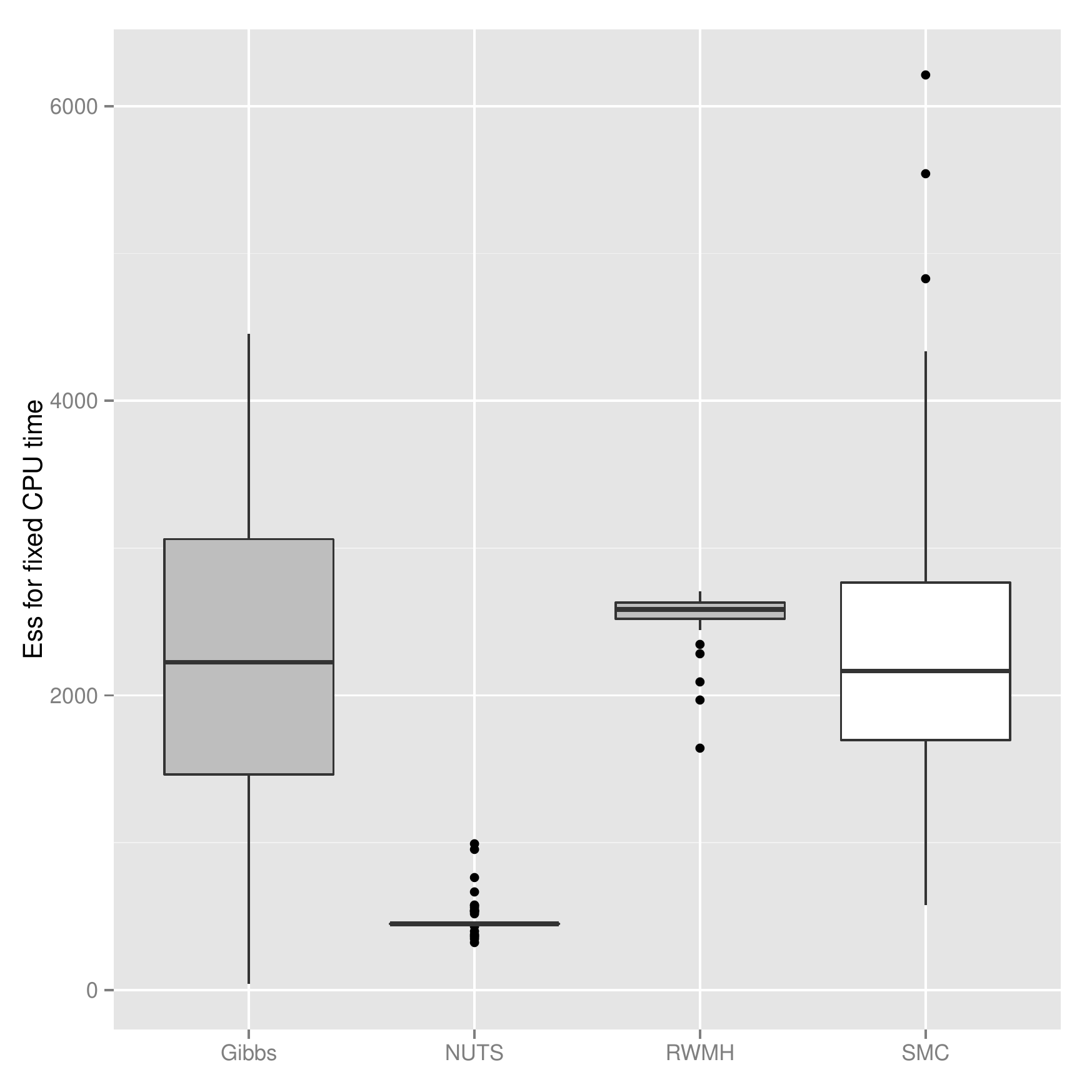}\protect\includegraphics[scale=0.35]{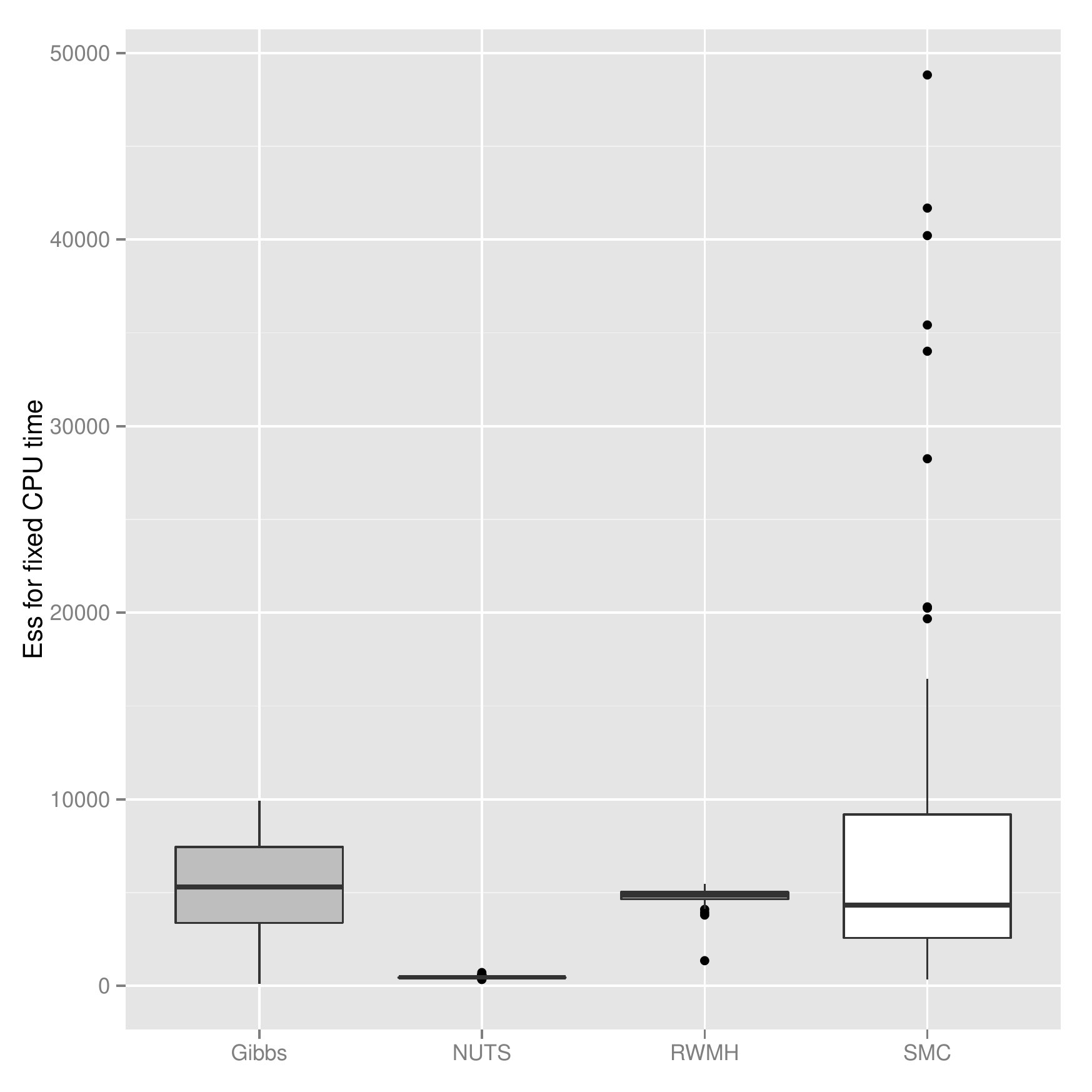}\protect
\par\end{centering}

}

\subfloat[Sonar]{\protect\begin{centering}
\protect\includegraphics[scale=0.35]{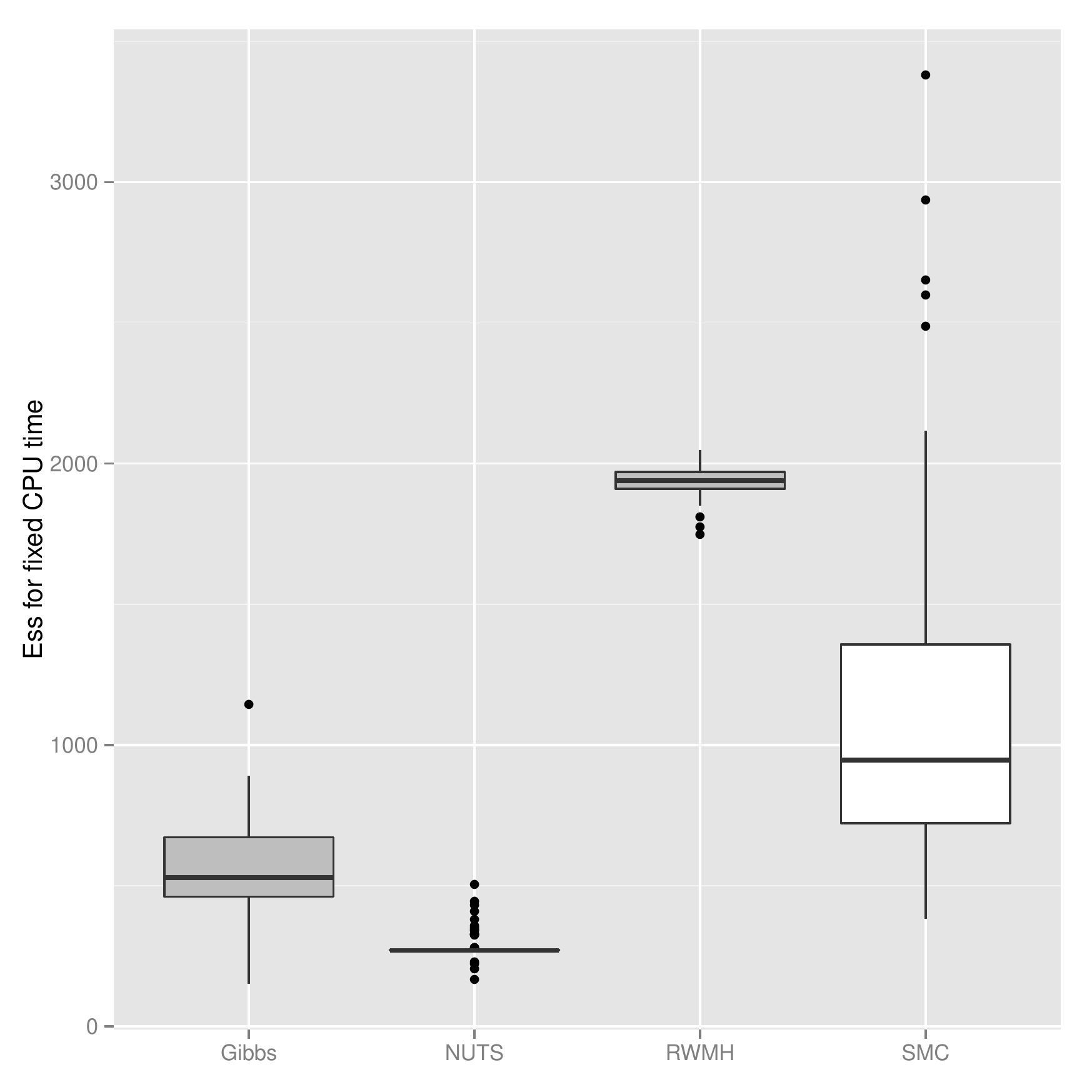}\protect\includegraphics[scale=0.35]{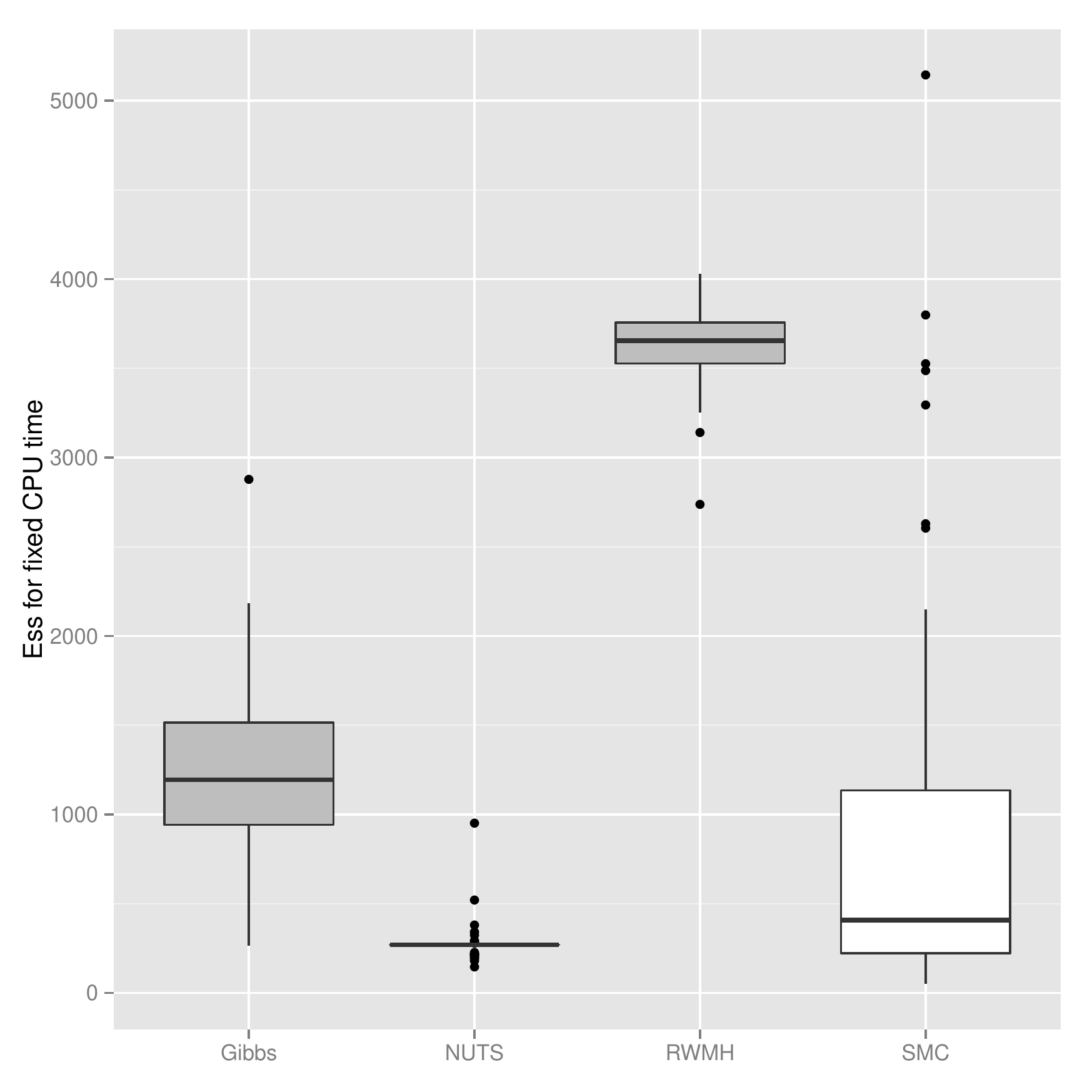}\protect
\par\end{centering}

}

\subfloat[DNA]{\protect\begin{centering}
\protect\includegraphics[scale=0.35]{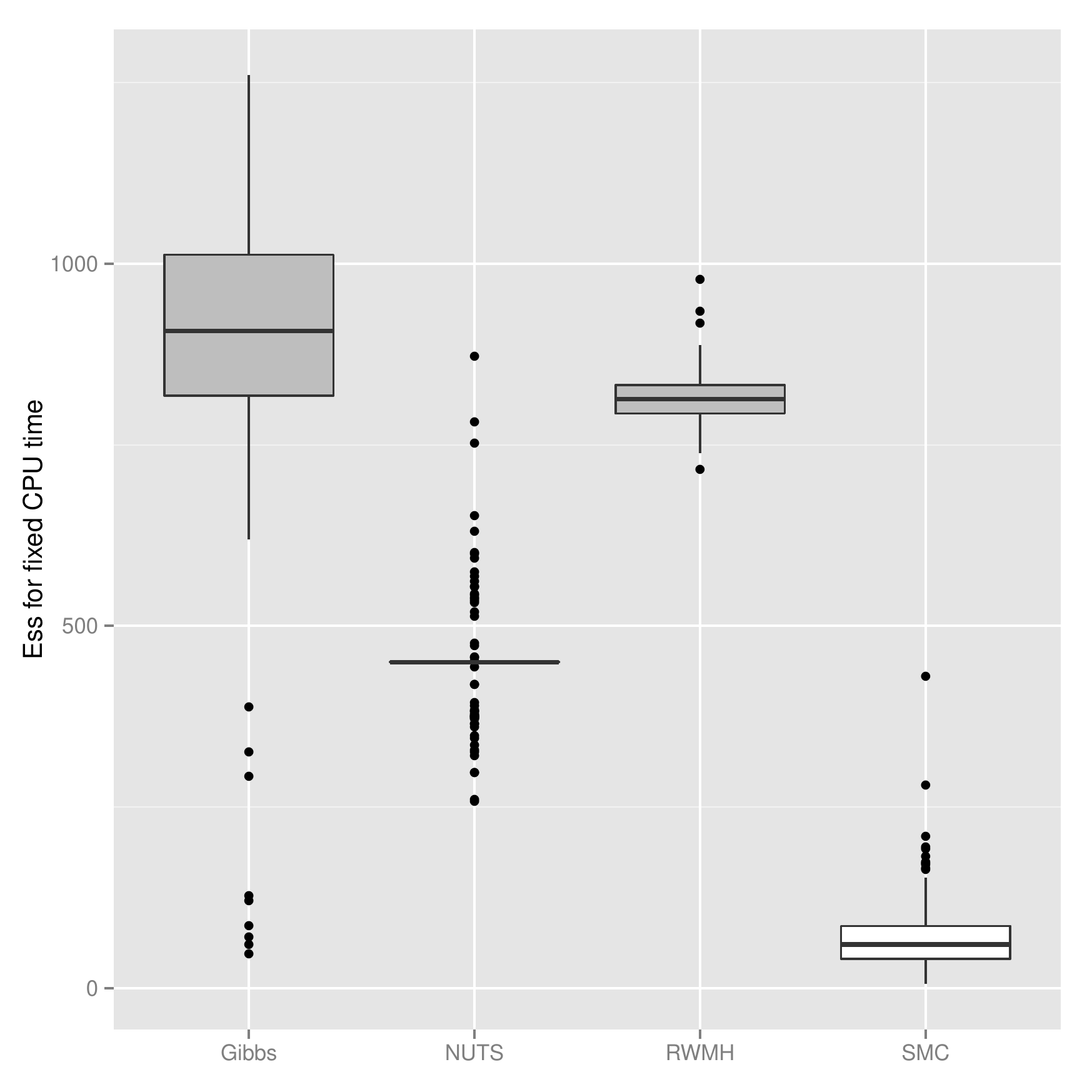}\protect\includegraphics[scale=0.35]{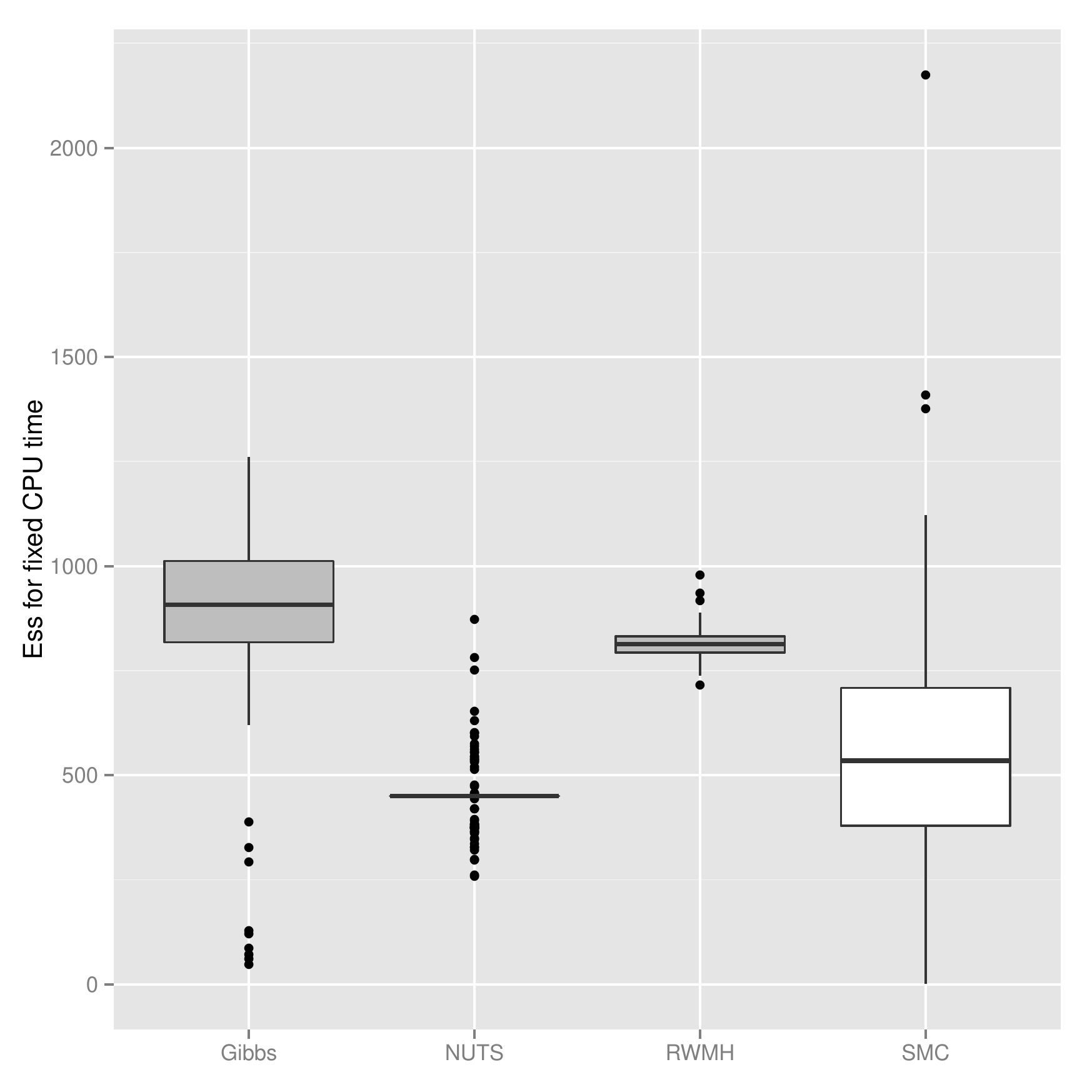}\protect
\par\end{centering}

}\protect\caption{\label{fig:ESS_big}Effective sample size for a fixed CPU time for
sampling-based algorithms: posterior expectations (left), and posterior
variances (right) for datasets (from top to bottom): Musk, Sonar,
and ADN}
\end{figure}

Figure \ref{fig:ESS_big} reports the ``effective sample size''
of the output of these algorithms when run for the same fixed CPU
time (corresponding to $5\times10^{5}$ iterations of RWHM), for the
$p$ posterior expectations (left panels), and the $p$ posterior
variances (right panels); here ``effective sample size'' is simply
the posterior variance divided by the MSE of the estimate (across
50 independent runs of the same algorithm). 

No algorithm seems to vastly outperform the others consistently across
the three datasets. If anything, RWMH seems to show consistently best
or second best performance. 

Still, these results offer the following insights. Again, we see that
Gibbs sampling, despite being a specialised algorithm, does not outperform
significantly more generic algorithms. Recall that the probit/Gaussian
scenario is very favourable to Gibbs sampling; in other scenarios
(results not shown), Gibbs is strongly dominated by other algorithms. 

More surprisingly, RWHM still performs well despite the high dimension.
In addition, RHHM seems more robust than SMC to an imperfect calibration;
see the DNA example, where the error of the EP approximation is greater. 

On the other hand, SMC is more amenable to parallelisation, hence
on a parallel architecture, SMC would be likely to outperform the
other approaches.

\section{Variable selection\label{sec:Variable-selection}}

We discuss in this section the implications of our findings on variable
selection. The standard way to formalise variable selection is to
introduce as a parameter the binary vector $\bgamma\in\left\{ 0,1\right\} ^{p}$,
and to define the likelihood 

\[
p(\data|\bbeta,\bgamma)=\prod_{i=1}^{\ndata}F(y_{i}\bbeta_{\bgamma}^{T}\bx_{\bgamma,i})
\]
where $\bbeta_{\bgamma}$ (resp. $\bx_{\bgamma,i}$) is the vector
of length $\left|\bgamma\right|$ that one obtains by excluding from
$\bbeta$ (resp. $\bx_{i}$) the components $j$ such that $\gamma_{j}=0$.
Several priors may be considered for this problem \citep{Chipman:practicalmodelselection},
but for simplicity, we will take $p(\bbeta,\bgamma)=p(\bbeta)p(\bgamma)$
where $p(\bbeta)$ is either the Cauchy prior or the Gaussian prior
discussed in Section \ref{sub:Likelihood,-prior}, and $p(\bgamma)$
is the uniform distribution with respect to the set $\left\{ 0,1\right\} ^{p}$,
$p(\bgamma)=2^{-p}$. 

Computationally, variable selection is more challenging than parameter
estimation, because the posterior $p(\bbeta,\bgamma|\data)$ is a
mixture of discrete and continuous components. If $p$ is small, one
may simply perform a complete enumeration: for all the $2^{p}$ possible
values of $\bgamma$, approximate $p(\data|\bgamma)$ using e.g. importance
sampling. If $p$ is large, one may adapt the approach of \citet{schafer2011sequential},
as described in the next sections.

\subsection{SMC algorithm of \citet{schafer2011sequential}}

In linear regression, $y_{i}=\bbeta_{\bgamma}^{T}\bx_{\bgamma,i}+\varepsilon_{i}$,
$\varepsilon_{i}\sim\Norm_{1}(0,\sigma^{2})$, the marginal likelihood
$p(\data|\bgamma)$ is available in close form (for a certain class
of priors). \citet{schafer2011sequential} use this property to construct
a tempering SMC sampler, which transitions from the prior $p(\bgamma)$
to the posterior $p(\bgamma|\data)$, through the tempering sequence
$\pi_{t}(\bgamma)\propto p(\bgamma)p(\data|\bgamma)^{\delta_{t}}$,
with $\delta_{t}$ growing from $0$ to $1$. This algorithm has the
same structure as Algorithm \ref{alg:tempering-SMC} (with the obvious
replacements of the $\bbeta$'s by $\bgamma$'s and so on.) The only
difference is the MCMC step used to diversify the particles after
resampling. Instead of a random walk step (which would be ill-defined
on a discrete space), \citet{schafer2011sequential} use a Metropolis
step based on an independent proposal, constructed from a sequence
of nested logistic regressions: proposal for first component $\gamma_{1}$
is Bernoulli, proposal for second component $\gamma_{2}$, conditional
on $\gamma_{1}$, corresponds to a logistic regression with $\gamma_{1}$
and an intercept as covariates, and so on. The parameters of these
$p$ successive regressions are simply estimated from the current
particle system. \citet{schafer2011sequential} show that their algorithm
significantly outperform several MCMC samplers on datasets with more
than $100$ covariates.

\subsection{Adaptation to binary regression}

For binary regression models, $p(\data|\bgamma)$ is intractable,
so the approach of \citet{schafer2011sequential} cannot be applied
directly. On the other hand, we have seen that (a) both Laplace and
EP may provide a fast approximation of the evidence $p(\data|\bgamma)$;
and (b) both importance sampling and the tempering SMC algorithm may
provide an \emph{unbiased }estimator of $p(\data|\bgamma)$. 

Based on these remarks, \citet{schafer:thesis} in his PhD thesis
considered the following extension of the SMC algorithm of \citet{schafer2011sequential}:
in the sequence $\pi_{t}(\bgamma)\propto p(\bgamma)p(\data|\bgamma)^{\delta_{t}}$,
the intractable quantity $p(\data|\bgamma)$ is simply replaced by
an unbiased estimator (obtained with importance sampling and the Gaussian
proposal corresponding to Laplace). The corresponding algorithm remains
valid, thanks to pseudo-marginal arguments \citep[see e.g.][]{Andrieu2009}.
Specifically, one may re-interpret the resulting algorithm as a SMC
algorithm for a sequence of distribution of an extended space, such
that marginal in $\bgamma$ is exactly the posterior $p(\data|\bgamma)$
at time $t=T$. In fact, it may be seen as a particular variant of
the SMC$^{2}$ algorithm of \citet{smc2}.

\subsection{Numerical illustration}

\begin{figure}
\subfloat[Gibbs]{\protect\includegraphics[scale=0.35]{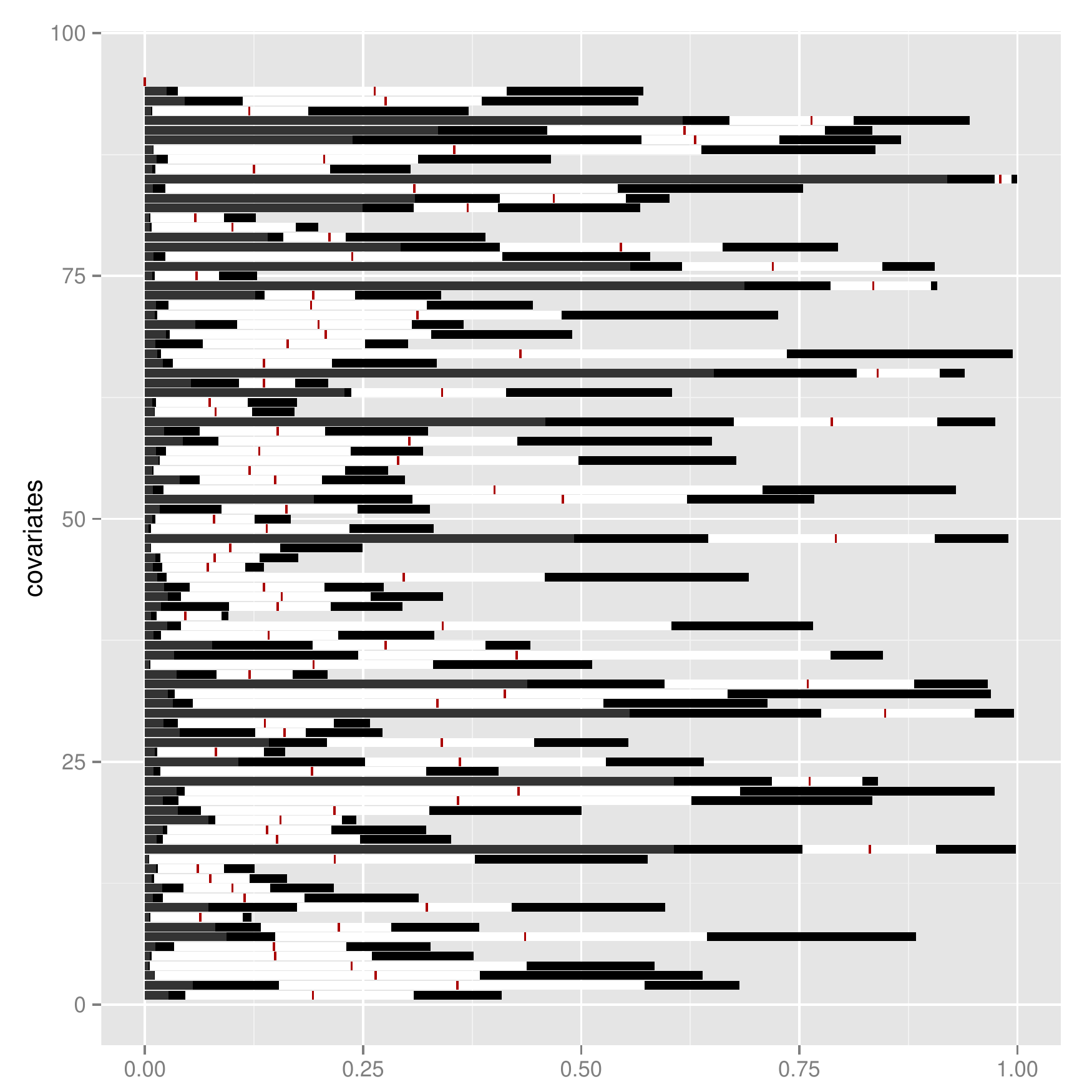}

}\subfloat[SMC]{\protect\includegraphics[scale=0.35]{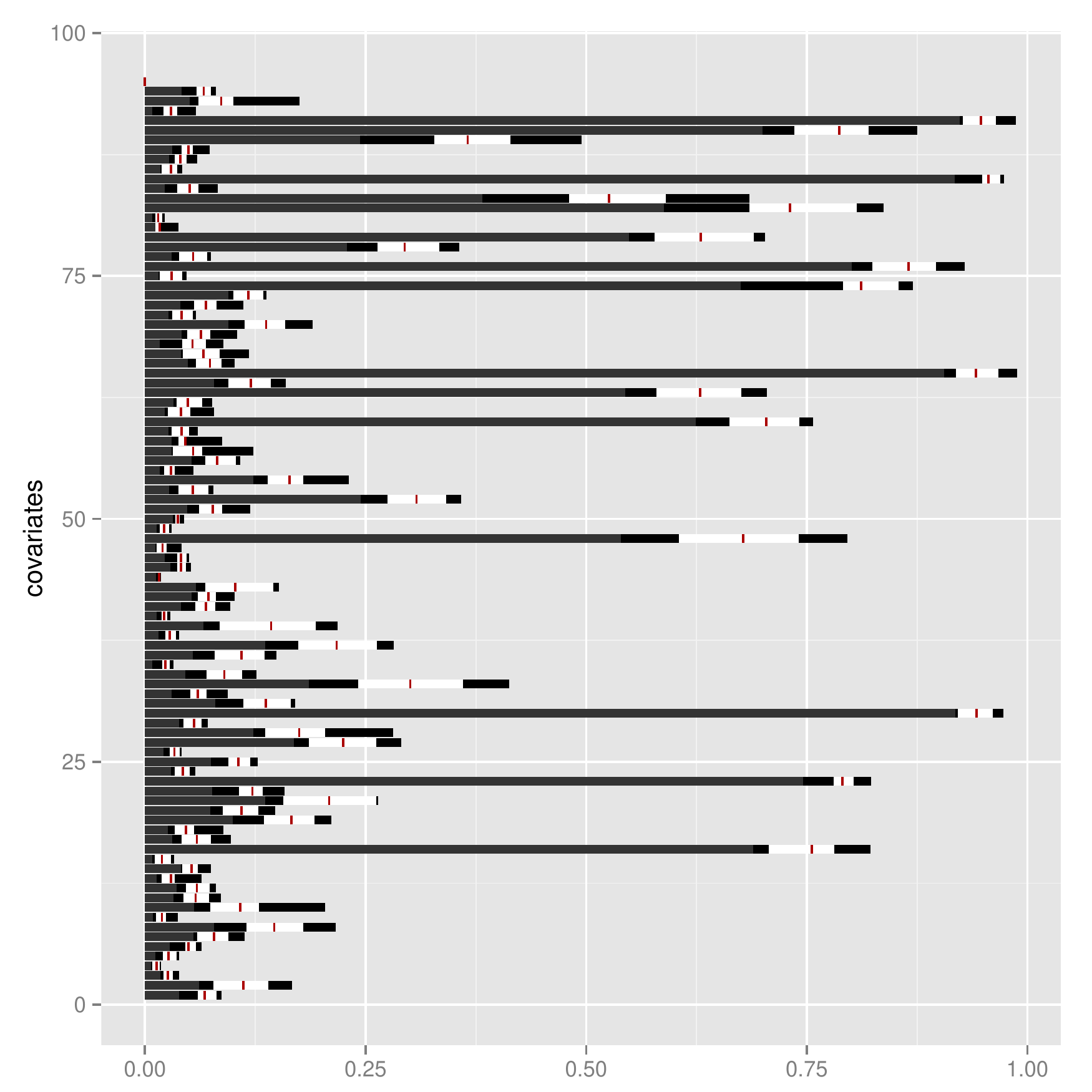}

}\protect\caption{\label{fig:varsel}Variation of estimated inclusion probabilities
$p(\gamma_{j}=1|\mathcal{D})$ over 50 runs for the $p$ covariates
of Musk dataset: median (red line), 80\% confidence interval (white
box); the black-box extends until the maximum value. }
\end{figure}

We now compare the proposed SMC approach with the Gibbs sampler of
\citet{Holmes2006} for sampling from $p(\bbeta,\bgamma|\mathcal{D})$,
on the Musk dataset. Both algorithms were given the same CPU budget
(15 minutes), and were run 50 times; see Figure \ref{fig:varsel}.
Clearly, the SMC sampler provides more reliable estimates of the inclusion
probabilities $p(\gamma_{j}=1|\mathcal{D})$ on such a big dataset.
See also the PhD dissertation of \citet{schafer:thesis} for results
consistent with those, on other datasets, and when comparing to the
adaptive reversible jump sampler of \citet{MR3173739}. 

\subsection{Spike and slab}

We also note in passing that a different approach to the variable
selection problem is to assign a spike and slab prior to $\bbeta$
\citep{George1993}: 

\[
p(\bbeta)=\prod_{j=1}^{p}\left\{ \lambda\Norm_{1}(\beta_{j};0,v_{0}^{2})+(1-\lambda)\Norm_{1}(\beta_{j};0,v_{1}^{2})\right\} ,\quad v_{0}^{2}\ll v_{1}^{2}
\]
where $\lambda\in(0,1)$, $v_{0}^{2}$ and $v_{1}^{2}$ are fixed
hyper-parameters. This prior generates a continuous posterior (without
point masses at $\beta_{j}=0$), which is easier to sample from than
the discrete-continuous mixture obtained in the standard formulation
of Bayesian variable selection. It would be interesting to see to
which extent our discussion and findings extend to this particular
type of posteriors; see for instance \citet{MR3104499} for how to
deal with such priors in EP.

\section{Conclusion and extensions\label{sec:Conclusion-and-extensions}}

\subsection{Our main messages to users}

Our first and perhaps most important message to end users is that
Bayesian computation (for binary regression) is now sufficiently fast
for routine use: if the right approach is used, results may be obtained
near instantly on a standard computer, at least on simple datasets. 

Concretely, as far as binary regression is concerned, our main recommendation
is to always use EP. It is very fast, and its approximation error
is negligible in most cases (for such models). EP requires some expertise
to implement, but the second author will release shortly a R package
that computes the EP approximation for any logit or probit model.
The only drawback of EP is the current lack of theoretical support.
We learnt however while finishing this manuscript that Simon Barthelmé
and Guillaume Dehaene (personal communication) established that the
error rate of EP is $\OO(\ndata^{-2})$ in certain models (where $\ndata$
is the sample size). This seems to explain why EP often performs so
well. 

In case one wishes to assess the EP error, by running in a second
step some exact algorithm, we would recommend to use the SMC approach
outlined in Section \ref{sub:Sequential-Monte-Carlo} (i.e. with initial
particles simulated from the EP approximation). Often, this SMC sampler
will reduce to a single importance sampling step, and will perform
extremely well. Even when it does not, it should provide decent performance,
especially if run on (and implemented for) a parallel architecture.
Alternatively, on a single-core machine, random walk Metropolis is
particularly simple to implement, and performs surprisingly well on
high-dimensional data (when properly calibrated using EP).

\subsection{Our main message to Bayesian computation experts}

Our main message to Bayesian computation scientists was already in
the title of this paper: leave Pima Indians alone, and more generally,
let's all refrain from now on from using datasets and models that
are too simple to serve as a reasonable benchmark. 

To elaborate, let's distinguish between specialised algorithms and
generic algorithms. 

For algorithms specialised to a given model and a given prior (i.e.
Gibbs samplers), the choice of a ``benchmark'' reduces to the choice
of a dataset. It seems unfortunate that such algorithms are often
showcased on small datasets (20 covariates or less), for which simpler,
more generic methods perform much better. As a matter of fact, we
saw in our simulations that even for bigger datasets Gibbs sampling
does not seem to offer better performance than generic methods.

For generic algorithms (Metropolis, HMC, and so on), the choice of
a benchmark amounts to the choice of a target distribution. A common
practice in papers proposing some novel algorithm for Bayesian computation
is to compare that algorithm with a Gibbs sampler on a binary regression
posterior for a small dataset. Again, we see from our numerical study
that this benchmark is of of limited interest, and may not be more
informative than a Gaussian target of the same dimension. If one wishes
to stick with binary regression, then datasets with more than 100
covariates should be used, and numerical comparisons should include
at least a properly calibrated random walk Metropolis sampler.

\subsection{Big data and the $p^{3}$ frontier}

Several recent papers \citep{wang2013parallelizing,ConsensusMonteCarlo,bardenet2015markov}
have approached the 'big data' problem in Bayesian computation by
focussing on the big $\ndata$ (many observations) scenario. In binary
regression, and possibly in similar models, the big $p$ problem (many
covariates) seems more critical, as the complexity of most the algorithms
we have discussed is $\OO(\ndata p^{3})$. Indeed, we do not believe
that any of the methods discussed in this paper is practical for $p\gg1000$.
The large $p$ problem may be therefore the current frontier of Bayesian
computation for binary regression. 

Perhaps one way to address the large $p$ problem is to make stronger
approximations; for instance by using EP with an approximation family
of sparse Gaussians. Alternatively, one may use a variable selection
prior that forbids that the number of active covariates is larger
than a certain threshold.

\subsection{Generalising to other models}

We suspect some of our findings may apply more generally to other
models (such as certain generalised linear models), but, of course,
further study is required to assess this statement. 

On the other hand, there are two aspects of our study which we recommend
to consider more generally when studying other models: parallelisation,
and taking into account the availability of fast approximations. The
former has already been discussed. Regarding the latter, binary regression
models are certainly not the only models such that some fast approximations
may be obtained, whether through Laplace, INLA, Variational Bayes,
or EP. And using this approximation to calibrate sampling-based algorithms
(Hastings-Metropolis, HMC, SMC, and so on)  will often have a dramatic
impact on the relative performance of these algorithms. Alternatively,
one may also discover in certain cases that these approximations are
sufficiently accurate to be used directly.

\section*{Acknowledgements}

We thank Håvard Rue for insightful comments. The first author is partially
funded by Labex ECODEC ANR - 11-LABEX-0047 grant from ANR (Agence
Nationale de la Recherche). 

\bibliographystyle{apalike}
\bibliography{complete}

\begin{thebibliography}{}

\bibitem[Albert and Chib, 1993]{Chib}
Albert, J.~H. and Chib, S. (1993).
\newblock Bayesian analysis of binary and polychotomous response data.
\newblock {\em J. Am. Statist. Assoc.}, 88(422):669--79.

\bibitem[Andrieu and Roberts, 2009]{Andrieu2009}
Andrieu, C. and Roberts, G. (2009).
\newblock {The pseudo-marginal approach for efficient Monte Carlo
  computations}.
\newblock {\em The Annals of Statistics}, 37(2):697--725.

\bibitem[Andrieu and Thoms, 2008]{andrieu2008tutorial}
Andrieu, C. and Thoms, J. (2008).
\newblock {A tutorial on adaptive MCMC}.
\newblock {\em Statist. Comput.}, 18(4):343--373.

\bibitem[Bardenet et~al., 2015]{bardenet2015markov}
Bardenet, R., Doucet, A., and Holmes, C. (2015).
\newblock {On Markov chain Monte Carlo methods for tall data}.
\newblock {\em arXiv preprint arXiv:1505.02827}.

\bibitem[Beskos et~al., 2013]{Beskos2013}
Beskos, A., Pillai, N., Roberts, G., Sanz-Serna, J.-M., and Stuart, A. (2013).
\newblock {Optimal tuning of the hybrid Monte Carlo algorithm}.
\newblock {\em Bernoulli}, 19(5A):1501--1534.

\bibitem[Bishop, 2006]{Bishop:book}
Bishop, C. (2006).
\newblock {\em {Pattern recognition and machine learning}}.
\newblock Springer New York.

\bibitem[Chipman et~al., 2001]{Chipman:practicalmodelselection}
Chipman, H., George, E.~I., and McCulloch, R.~E. (2001).
\newblock {\em The practical implementation of Bayesian model selection}, pages
  65--134.

\bibitem[Chopin, 2002]{Chopin:IBIS}
Chopin, N. (2002).
\newblock A sequential particle filter for static models.
\newblock {\em Biometrika}, 89:539--552.

\bibitem[Chopin, 2011]{chopin2011fast}
Chopin, N. (2011).
\newblock {Fast simulation of truncated Gaussian distributions}.
\newblock {\em Statist. Comput.}, 21(2):275--288.

\bibitem[Chopin et~al., 2013]{smc2}
Chopin, N., Jacob, P., and Papaspiliopoulos, O. (2013).
\newblock {SMC$^2$: A sequential Monte Carlo algorithm with particle Markov
  chain Monte Carlo updates}.
\newblock {\em J. R. Statist. Soc. B}, 75(3):397--426.

\bibitem[Consonni and Marin, 2007]{consonni2007mean}
Consonni, G. and Marin, J. (2007).
\newblock Mean-field variational approximate {B}ayesian inference for latent
  variable models.
\newblock {\em Comput. Stat. Data Anal.}, 52(2):790--798.

\bibitem[Cook, 2014]{Cook:time}
Cook, J.~D. (2014).
\newblock Time exchange rate.
\newblock {\em The Endeavour (blog)}.

\bibitem[Del~Moral et~al., 2006]{DelDouJas:SMC}
Del~Moral, P., Doucet, A., and Jasra, A. (2006).
\newblock {Sequential Monte Carlo samplers}.
\newblock {\em Journal of the Royal Statistical Society: Series B (Statistical
  Methodology)}, 68(3):411--436.

\bibitem[Dempster et~al., 1977]{DemLaiRub}
Dempster, A.~P., Laird, N.~M., and Rubin, D.~B. (1977).
\newblock {Maximum likelihood from incomplete data via the EM algorithm}.
\newblock {\em J. R. Statist. Soc. B}, 39:1--38.

\bibitem[Duane et~al., 1987]{Duane1987}
Duane, S., Kennedy, A., Pendleton, B.~J., and Roweth, D. (1987).
\newblock {Hybrid Monte Carlo}.
\newblock {\em Physics Letters B}, 195(2):216--222.

\bibitem[Faes et~al., 2011]{Faes2011}
Faes, C., Ormerod, J.~T., and Wand, M.~P. (2011).
\newblock Variational {B}ayesian inference for parametric and nonparametric
  regression with missing data.
\newblock {\em Journal of the American Statistical Association},
  106(495):959--971.

\bibitem[Firth, 1993]{firth1993bias}
Firth, D. (1993).
\newblock Bias reduction of maximum likelihood estimates.
\newblock {\em Biometrika}, 80(1):27--38.

\bibitem[Fr\"uhwirth-Schnatter and Fr\"uhwirth, 2009]{FrhwirthSchnatter2009}
Fr\"uhwirth-Schnatter, S. and Fr\"uhwirth, R. (2009).
\newblock {Data augmentation and MCMC for Binary and multinomial logit models}.
\newblock In {\em Statistical Modelling and Regression Structures}, pages
  111--132. Physica-Verlag HD.

\bibitem[Gelman and Hill, 2006]{gelman2006data}
Gelman, A. and Hill, J. (2006).
\newblock {\em Data analysis using regression and multilevel/hierarchical
  models}.
\newblock Cambridge University Press.

\bibitem[Gelman et~al., 2008]{MR2655663}
Gelman, A., Jakulin, A., Pittau, M.~G., and Su, Y.-S. (2008).
\newblock A weakly informative default prior distribution for logistic and
  other regression models.
\newblock {\em Ann. Appl. Stats.}, 2(4):1360--1383.

\bibitem[George and McCulloch, 1993]{George1993}
George, E.~I. and McCulloch, R.~E. (1993).
\newblock {Variable Selection via Gibbs Sampling}.
\newblock {\em J. Am. Statist. Assoc.}, 88(423):881--889.

\bibitem[Girolami and Calderhead, 2011]{girolami2011riemann}
Girolami, M. and Calderhead, B. (2011).
\newblock Riemann manifold {L}angevin and {H}amiltonian {M}onte {C}arlo
  methods.
\newblock {\em J. R. Statist. Soc. B}, 73(2):123--214.

\bibitem[Gramacy and Polson, 2012]{Gramacy2012}
Gramacy, R.~B. and Polson, N.~G. (2012).
\newblock Simulation-based regularized logistic regression.
\newblock {\em Bayesian Anal.}, 7(3):567--590.

\bibitem[Hern{\'a}ndez-Lobato et~al., 2013]{MR3104499}
Hern{\'a}ndez-Lobato, D., Hern{\'a}ndez-Lobato, J.~M., and Dupont, P. (2013).
\newblock Generalized spike-and-slab priors for {B}ayesian group feature
  selection using expectation propagation.
\newblock {\em J. Mach. Learn. Res.}, 14:1891--1945.

\bibitem[Hoffman and Gelman, 2013]{NUTS}
Hoffman, M. and Gelman, A. (2013).
\newblock The no-{U}-turn sampler: Adaptively setting path lengths in
  {H}amiltonian monte carlo.
\newblock {\em J. Machine Learning Research}, page (in press).

\bibitem[Holmes and Held, 2006]{Holmes2006}
Holmes, C.~C. and Held, L. (2006).
\newblock Bayesian auxiliary variable models for binary and multinomial
  regression.
\newblock {\em Bayesian Anal.}, 1(1):145--168.

\bibitem[H{\"o}rmann and Leydold, 2005]{hormann2005quasi}
H{\"o}rmann, W. and Leydold, J. (2005).
\newblock Quasi importance sampling.
\newblock Technical report.

\bibitem[Jacob et~al., 2011]{Jacob2011}
Jacob, P., Robert, C.~P., and Smith, M.~H. (2011).
\newblock {Using Parallel Computation to Improve Independent
  Metropolis$\textendash$Hastings Based Estimation}.
\newblock {\em J. Comput. Graph. Statist.}, 20(3):616--635.

\bibitem[Jasra et~al., 2011]{jasrainference}
Jasra, A., Stephens, D., A.~Doucet, A., and Tsagaris, T. (2011).
\newblock Inference for {L}\'evy driven stochastic volatility models via
  {S}equential {M}onte {C}arlo.
\newblock {\em Scand. J. of Statist.}, 38(1).

\bibitem[Kab{\'a}n, 2007]{kaban2007bayesian}
Kab{\'a}n, A. (2007).
\newblock {On Bayesian classification with Laplace priors}.
\newblock {\em Pattern Recognition Letters}, 28(10):1271--1282.

\bibitem[Kong et~al., 1994]{Kong1994}
Kong, A., Liu, J.~S., and Wong, W.~H. (1994).
\newblock {Sequential imputation and Bayesian missing data problems}.
\newblock {\em J. Am. Statist. Assoc.}, 89:278--288.

\bibitem[Lamnisos et~al., 2013]{MR3173739}
Lamnisos, D., Griffin, J.~E., and Steel, M. F.~J. (2013).
\newblock Adaptive {M}onte {C}arlo for {B}ayesian variable selection in
  regression models.
\newblock {\em J. Comput. Graph. Statist.}, 22(3):729--748.

\bibitem[Lee et~al., 2010]{Lee2010}
Lee, A., Yau, C., Giles, M.~B., Doucet, A., and Holmes, C.~C. (2010).
\newblock On the utility of graphics cards to perform massively parallel
  simulation of advanced {M}onte {C}arlo methods.
\newblock {\em J. Comput. Graph. Statist.}, 19(4):769--789.

\bibitem[Lemieux, 2009]{Lemieux:MCandQMCSampling}
Lemieux, C. (2009).
\newblock {\em {Monte Carlo and Quasi-Monte Carlo Sampling (Springer Series in
  Statistics)}}.
\newblock Springer.

\bibitem[Minka, 2001]{minka2001expectation}
Minka, T. (2001).
\newblock {Expectation Propagation for approximate Bayesian inference}.
\newblock {\em Proceedings of Uncertainty in Artificial Intelligence},
  17:362--369.

\bibitem[Neal, 2001]{Neal:AIS}
Neal, R.~M. (2001).
\newblock Annealed importance sampling.
\newblock {\em Statist. Comput.}, 11:125--139.

\bibitem[Neal, 2010]{Neal2010HMC}
Neal, R.~M. (2010).
\newblock {MCMC} using {H}amiltonian dynamics.
\newblock In Brooks, S., Gelman, A., Jones, G.~L., and Meng, X.-L., editors,
  {\em Handbook of Markov Chain Monte Carlo}, pages 113--162. Chapman \& Hall /
  CRC Press.

\bibitem[Nickisch and Rasmussen,
  2008]{NickishRasmussen:ApproxGaussianProcClass}
Nickisch, H. and Rasmussen, C. (2008).
\newblock {Approximations for Binary Gaussian Process Classification}.
\newblock {\em J. Machine Learning Research}, 9(10):2035--2078.

\bibitem[Polson et~al., 2013]{Polson2013}
Polson, N.~G., Scott, J.~G., and Windle, J. (2013).
\newblock Bayesian inference for logistic models using
  p\'olya$\textendash$gamma latent variables.
\newblock {\em Journal of the American Statistical Association},
  108(504):1339--1349.

\bibitem[Press et~al., 2007]{NumericalRecipes3rdEd}
Press, W.~H., Teukolsky, S.~A., Vetterling, W.~T., and Flannery, B.~P. (2007).
\newblock {\em Numerical Recipes: The Art of Scientific Computing}.
\newblock Cambridge University Press.

\bibitem[Ridgway, 2014]{ridgway2014computation}
Ridgway, J. (2014).
\newblock Computation of {G}aussian orthant probabilities in high dimension.
\newblock {\em arXiv preprint arXiv:1411.1314}.

\bibitem[Robert and Casella, 2004]{RobCas}
Robert, C.~P. and Casella, G. (2004).
\newblock {\em {M}onte {C}arlo Statistical Methods, 2nd ed.}
\newblock Springer-Verlag, New York.

\bibitem[Roberts and Rosenthal, 2001]{RobertsRosenthal:OptimalScalingMH}
Roberts, G.~O. and Rosenthal, J.~S. (2001).
\newblock {Optimal scaling for various Metropolis-Hastings algorithms}.
\newblock {\em Statist. Science}, 16(4):351--367.

\bibitem[Rue et~al., 2009]{rue2009approximate}
Rue, H., Martino, S., and Chopin, N. (2009).
\newblock {Approximate Bayesian inference for latent Gaussian models by using
  integrated nested Laplace approximations}.
\newblock {\em J. R. Statist. Soc. B}, 71(2):319--392.

\bibitem[Sch\"afer, 2012]{schafer:thesis}
Sch\"afer, C. (2012).
\newblock {\em Monte Carlo methods for sampling high-dimensional binary
  vectors}.
\newblock PhD thesis, Universit\'e Paris Dauphine.

\bibitem[Sch{\"a}fer and Chopin, 2011]{schafer2011sequential}
Sch{\"a}fer, C. and Chopin, N. (2011).
\newblock Sequential monte carlo on large binary sampling spaces.
\newblock {\em Statistics and Computing}, pages 1--22.

\bibitem[Scott et~al., 2013]{ConsensusMonteCarlo}
Scott, S.~L., Blocker, A.~W., and Bonassi, F.~V. (2013).
\newblock Bayes and big data: The consensus monte carlo algorithm.
\newblock In {\em Bayes 250}.

\bibitem[Seeger, 2005]{Seeger:EPExpFam}
Seeger, M. (2005).
\newblock {Expectation Propagation for Exponential Families}.
\newblock Technical report, Univ. California Berkeley.

\bibitem[Shahbaba et~al., 2011]{shahbaba2011split}
Shahbaba, B., Lan, S., Johnson, W.~O., and Neal, R.~M. (2011).
\newblock {Split Hamiltonian Monte Carlo}.
\newblock {\em Statist. Comput.}, pages 1--11.

\bibitem[Skilling, 2006]{SkillingNested}
Skilling, J. (2006).
\newblock {Nested sampling for general Bayesian computation}.
\newblock {\em Bayesian Analysis}, 1(4):833--860.

\bibitem[Suchard et~al., 2010]{Suchard2010}
Suchard, M.~A., Wang, Q., Chan, C., Frelinger, J., Cron, A., and West, M.
  (2010).
\newblock Understanding {GPU} programming for statistical computation: Studies
  in massively parallel massive mixtures.
\newblock {\em J. Comput. Graph. Statist.}, 19(2):419--438.

\bibitem[Tierney and Kadane, 1986]{tierney1986accurate}
Tierney, L. and Kadane, J.~B. (1986).
\newblock Accurate approximations for posterior moments and marginal densities.
\newblock {\em J. Am. Statist. Assoc.}, 81(393):82--86.

\bibitem[Tierney et~al., 1989]{Kass}
Tierney, L., Kass, R.~E., and Kadane, J.~B. (1989).
\newblock Fully exponential {L}aplace approximations to expectations and
  variances of non-positive functions.
\newblock {\em J. Am. Statist. Assoc.}, 84:710--716.

\bibitem[van Gerven et~al., 2010]{vanGerven2010}
van Gerven, M.~A., Cseke, B., de~Lange, F.~P., and Heskes, T. (2010).
\newblock Efficient {B}ayesian multivariate {fMRI} analysis using a sparsifying
  spatio-temporal prior.
\newblock {\em {NeuroImage}}, 50(1):150--161.

\bibitem[Wang and Dunson, 2013]{wang2013parallelizing}
Wang, X. and Dunson, D.~B. (2013).
\newblock {Parallelizing MCMC via Weierstrass sampler}.
\newblock {\em arXiv preprint arXiv:1312.4605}.

\end{thebibliography}

\end{document}